\documentclass[aps,pra,twocolumn,superscriptaddress,nofootinbib,longbibliography]{revtex4-2}

\usepackage{amsmath,amssymb}
\usepackage{graphicx}
\usepackage{bm}
\usepackage{xcolor}
\usepackage{hyperref}

\begin{document}

\title{Quantum-to-classical transition and the emergence of trajectory level Darwinism with
measurements distributed in time: a path integral approach}

\author{Harsh Arora}
\email{harsharora@iisc.ac.in}
\affiliation{Centre for Nano Science and Engineering, Indian Institute of Science,
Bengaluru, India 560094}

\author{Bishal Kumar Das}
\email{bishal.9197@physics.iitm.ac.in}
\affiliation{Department of Physics, Indian Institute of Technology Madras,
Chennai, India 600036}
\affiliation{Center for Quantum Information, Communication and Computing,
Indian Institute of Technology Madras, Chennai, India 600036}

\author{Baladitya Suri}
\email{surib@iisc.ac.in}
\affiliation{Instrumentation and Applied Physics, Indian Institute of Science,
Bengaluru, India 560094}

\author{Vaibhav Madhok}
\email{madhok@physics.iitm.ac.in}
\affiliation{Department of Physics, Indian Institute of Technology Madras,
Chennai, India 600036}
\affiliation{Center for Quantum Information, Communication and Computing,
Indian Institute of Technology Madras, Chennai, India 600036}

\begin{abstract}
 We present a path-integral formulation of the emergence of classical dynamics that incorporates continuous measurements. Each Feynman path picks up a random phase due to measurements, and we give conditions under which the dominant contribution comes from the vicinity of the classical paths. This information proliferates across the environment, as in quantum Darwinism, via scattering of probes by the system. The scattering which decoheres a trajectory also heats it. This bounds how redundantly an individual classical trajectory can be recorded before back-action turns it into Brownian motion.

\end{abstract}

\maketitle

\section{Introduction}

The universe, as we know it, is quantum mechanical. There has not been a single
experimental violation of quantum theory. However, classical mechanics provides an
excellent description of the macroscopic world. While any macroscopic system is made up of
atoms, each of which does not obey Newton's laws, the system follows classical
trajectories governed by the same laws. How does classical mechanics arise from the
underlying quantum reality? A related question is the emergence of classical chaos from
the underlying unitary evolution of quantum mechanics. Classically chaotic trajectories,
which exhibit an exponential sensitivity to initial conditions, seem to arise from an
underlying unitary evolution dictated by the Schr\"odinger equation. The quantum world
does not show such an exponential sensitivity to initial conditions, as unitary evolution
causes the distance between two nearby quantum states to be a constant of
motion~\cite{haake1991quantum,asher_peres_chaos,peres1997quantum}. The reason for this apparent paradox is that
quantum states are analogous to probability distributions and not individual classical
trajectories. The Liouville and Koopman~\cite{koopman_pnas.17.5.315,von_neuman_1932,stepping_stone_hamiltonian}
formulation of classical mechanics preserves the overlap between two evolving probability
distributions. Nonetheless, it is the individually chaotic classical trajectories that are
observed in the macroscopic world and not probability distributions; therefore, the
question of their recovery from the unitarily evolving quantum state vectors is of
fundamental importance~\cite{TanmoyPRL}.

At first, it seems the correspondence principle addresses this question: just as one
recovers Galilean physics in the limit $c\to\infty$, one hopes to recover Newtonian
physics in the limit $\hbar\to0$. The two limits, however, are of very different
character. The limit $c\to\infty$ is regular: corrections are analytic in $1/c^2$,
relative errors remain uniformly of order $v^2/c^2$, and the time over which Galilean
predictions remain accurate to a given tolerance grows polynomially in $c$. The limit
$\hbar\to0$, by contrast, is singular~\cite{Berry2002}. Solutions of the Schr\"odinger
equation depend non-analytically on $\hbar$ (e.g.\ through phases $e^{iS/\hbar}$), and
the convergence to classical dynamics is not uniform in time. For any fixed $t$,
quantum predictions approach classical ones as $\hbar\to0$; for any fixed $\hbar$,
however, they eventually disagree completely as $t\to\infty$. The limits $\hbar\to0$
and $t\to\infty$ therefore do not commute. For chaotic systems, this crossover occurs
at the Ehrenfest time
\begin{equation}
  t_{Eh} \simeq \frac{1}{\lambda_L}\ln\!\left(\frac{S}{\hbar}\right),
\end{equation}
where $\lambda_L$ is the Lyapunov exponent and $S$ a characteristic classical action,
so that maintaining quantum--classical correspondence up to time $T$ requires
\begin{equation}
  \frac{S}{\hbar} \sim e^{\lambda_L T}.
\end{equation}
Where Galilean physics requires only a polynomially large $c$, Newtonian physics in a
chaotic system requires an exponentially large $S/\hbar$. Consequently, there is no
uniform criterion for what counts as ``macroscopic'': the required magnitude of
$S/\hbar$ depends on both the dynamics and the time scale of interest. Moreover, macroscopic quantum
superpositions are not ruled out by quantum theory. Yet, we do not see them.
While relativity shifts our perspective about space, time and frames of reference, quantum
mechanics denies reality to physical observables prior to
measurement~\cite{EPR,JSB}. Taking this argument to extreme limits, since the
entire universe, including the system, probe, and measurement apparatus, is quantum
mechanical, one cannot ascribe reality to anything. Therefore, quantum measurements, with
and even without invoking the mysterious ``collapse postulate,'' pose intriguing
conceptual and philosophical issues~\cite{EPR,bohr_1935,bohreinstein}. One way out of this
is to separate the universe into quantum and classical parts and rely on the latter as
accounting for our observable reality, with quantum measurements providing an entry ticket
into it~\cite{zurek_1991}.

The question of the quantum-to-classical transition is not to be confused with the
measurement problem (the ``collapse'' postulate) as mentioned above. Quantum theory gives
correct predictions for the outcome of experiments in a probabilistic sense. While the
density matrix formalism and the ensemble interpretation of quantum states provide a
statistical interpretation of quantum mechanics, in the real world we observe individual
trajectories rather than probabilities. Moreover, the quantum-to-classical transition
poses a challenge even for single quantum systems and one-shot instances of experiments.
After all, we observe individual trajectories of macroscopic objects, monitor, measure and
control single-shot experiments, and a statistical interpretation is more of a
book-keeping device rather than having an ontological reality of its own.

A natural formalism that bridges quantum mechanical coherence to actual trajectories is
the Feynman path integral formulation. In quantum theory the probability of an event which
can happen in several different ways is the absolute square of a sum of complex
contributions, the probability amplitudes, one from each alternative way. The probability
that a particle will be found to have a path in a region of space-time is the square of a
sum of complex contributions from all the paths in the region. The probability amplitude
from a single path is a complex number of unit magnitude whose (imaginary) phase is the
classical action (in units of $\hbar$) for the path in question. In other words, the
interference of all possible paths results in the determination of the final probability of
a path $x(t)$ in a given region. In the limit when the system action $S\gg\hbar$, the
dominant contribution comes from trajectories in the vicinity of the classical paths, which
are paths with least action (``extremal'' action), i.e.\ the first-order change in system
action for nearby paths is zero. Therefore, one would expect that for macroscopic systems,
it is the space-time region in the vicinity of the classical paths where the particle is
most likely to be found.

For a closed system, however, the stationary-phase argument holds only up to the Ehrenfest time.
In a chaotic macroscopic system an initially localized wave packet is stretched exponentially, and
coherent superpositions over macroscopic scales develop within the Ehrenfest time
~\cite{BermanZaslavsky1978,ZurekPaz1994,HabibShizumeZurek1998}. Since this time grows only logarithmically as $\hbar\to0$, it is short on everyday
timescales for typical chaotic systems, even though $\hbar$ is very small. A closed macroscopic
system therefore does not remain on a classical path indefinitely. The superpositions generated by
the dynamics have to be suppressed continually, and in all approaches to the quantum-to-classical
transition this is done by the environment monitoring the system. We are thus led to ask: what are
the dominant Feynman paths when environmental interactions, or measurements of the system via
probes, are included? In particular, under what measurement strength can one hope to actually
observe individual classical trajectories?

 In a related sense, one may
also seek the conditions under which the results from the framework of environment-induced
decoherence and the emergence of classicality at the level of density matrices can be
reconciled with the path integral framework. Therefore, in this work, we develop a
framework for the quantum-to-classical transition by considering a path integral
formulation of measurements distributed in time. This enables us to work at the level of
individual system trajectories and therefore retains the elegance of Feynman's original
approach as well as takes into account the role of the measurements.

Evolution of the system wave function conditioned on the measurement record reveals
classical motion for the quantum system if the effects of measurement-induced
localization, wave-packet spread due to the dynamics, and back-action noise balance each
other~\cite{TanmoyPRL}. Mathematically speaking, the evolution under
continuous coarse-grained measurements is given by a stochastic master equation that
suppresses probability amplitudes at positions away from the measured outcome by
``filtering'' the wave function ($\tilde x$): $\psi_m\propto\Upsilon(\tilde x-x)\psi(x)$.
This coarse-graining of the measurement gives rise to a stochastic evolution for the
measurement record:
\begin{equation}
r_{\rm meas}\sim r_{cl}+\alpha\zeta,
\label{eq:rmeas}
\end{equation}
where $\alpha$ sets the measurement imprecision and $\zeta$ is standard white noise. A smaller
$\alpha$ corresponds to stronger, more finely resolved monitoring.
 In the path integral formalism, the finite resolution of the measurement allows for
interference between trajectories, which singles out stationary paths as the ones with the
stationary phases, so that neighboring contributions interfere coherently. Therefore, in the limit $\hbar\to0$, the paths must be perturbed in
such a way that the phase remains stationary along them and neighboring contributions still add coherently. We show that, as an
alternative to the wave function collapse picture, the path integral framework with phase
scrambling (enforced by quantum complementarity) combined with the limit of continuous
measurements results in the measurement record picking up single classical trajectories.

\section{Heuristics}

\begin{figure}
    \centering
    \includegraphics[width=1\linewidth]{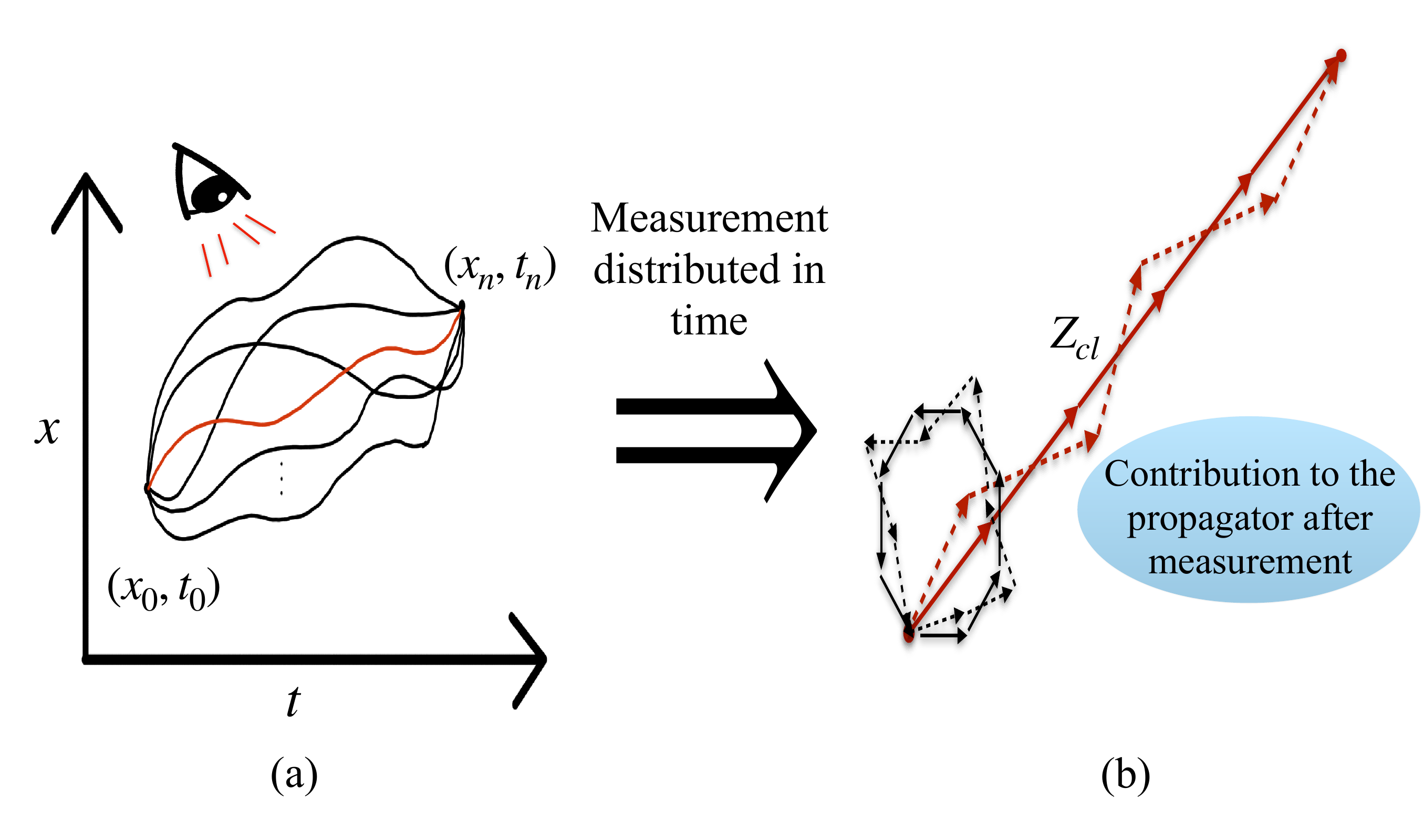}
    \caption{Weak measurements distributed in time might change the contribution from each path to the propagator; however, for macroscopic systems, they do not change the actual classical path: robustness of classicality in the presence of measurements. }
    \label{scrambling}
\end{figure}

The Feynman path integral representation for the transition probability amplitude
$\langle x_f|e^{-i\hat H(t_f-t_i)}|x_i\rangle$ works by assigning a probability amplitude
to each path that the system could take from $|x_i\rangle$ to $|x_f\rangle$. Therefore,
the overall amplitude is calculated by adding contributions of all paths:
\begin{equation}
K(x_f,t_f;x_i,t_i)\sim\sum_{\rm all\ paths}e^{\frac{iS[x(t)]}{\hbar}},
\label{eq:K-all}
\end{equation}
where $S[x(t)]$ is the action associated with a particular path. Each path can be
pictorially represented by a unit vector in the complex plane oriented in a particular
direction given by the phase $S[x(t)]/\hbar$; the sum over all paths is the addition of all
these vectors. Under the stationary phase approximation, the sum over all paths is
considerably simplified, since one only considers the contributions from paths in the
vicinity of the classical paths, leading to:
\begin{equation}
K(x_f,t_f;x_i,t_i)=\sum_{\rm classical\ paths}A_{cl}\,e^{\frac{iS_{cl}}{\hbar}},
\label{eq:K-cl}
\end{equation}
where $S_{cl}$ corresponds to the action evaluated along the classical paths. The above
approximation, which neglects contributions from non-classical paths since they add
destructively, becomes exact for systems in the macroscopic regime for which
$S_{cl}\gg\hbar$. For such systems, only a small fraction of paths with an action within a
narrow window $|\delta S|\lesssim\hbar$ of the action provide the dominant contribution to the total probability
amplitude Eq.~\eqref{eq:K-all}. Moreover, these paths are virtually indistinguishable from
the classical trajectory. Under measurements, each path can be regarded as acquiring
another phase $\theta_M$, as shown in Fig.~\ref{scrambling}. Depending on the strength
of the measurement, this phase can be a small perturbation (weak measurement) to the
original phase, or it can lead to a complete randomization of the original phase (strong
measurement). In the former case, as long as the measurement strength is approximately
within $|\delta S|\lesssim\hbar$, the paths in the vicinity of classical paths still contribute
coherently, and we can expect not only to recover individual classical trajectories but
also to observe them. In the case of a strong measurement, each path gets an additional
phase that is completely random. This is exactly the mechanism for decoherence via ``phase
randomization,'' where one recovers a classical picture of the quantum world at the level
of diagonal density matrices of the system state akin to a classical probability
distribution.

This phase randomization due to strong measurement causes rapid vanishing of the
off-diagonal elements of the system density matrix responsible for coherent quantum
phenomena. In other words,
$\sum_{\rm all\ paths}e^{\frac{iS[x(t)]}{\hbar}}\to\sum_{\rm all\ paths}e^{\frac{iS_M[x(t)]}{\hbar}}$,
where the modified action taking into account the measurement is
$S_M[x(t)]=S[x(t)]+\delta S$. Under the conditions $S_{cl}[x(t)]\gg\hbar$ and a
perturbation to the action $\delta S\approx\hbar$, the sum over paths becomes
$\sum_{\rm all\ paths}e^{\frac{iS_M[x(t)]}{\hbar}}\approx
\sum_{\rm all\ paths}e^{\frac{iS_{cl}[x(t)]+\delta S}{\hbar}}$, where $\delta S$ is a
random variable. Therefore, in the regime where the system action is macroscopic enough,
$S_{cl}[x(t)]\gg\hbar$, and under weak measurement in the sense above, one has a
quantum-to-classical transition of a system interacting with the environment or measured by
an external apparatus. In this way, the interference of all possible paths results in the
stationary paths as the dominant contributors. Further, the above expression becomes exact
in the limit $\hbar\to0$. Then, is it reasonable to expect that an attempt to continuously
measure an observable of the system will lead to classical results at the level of
individual trajectories?

The path integral formalism allows us to calculate quantum-mechanical amplitudes
corresponding to classical events, such as the probability amplitude that a particular path
is taken by the quantum system in phase space. The act of measurement necessarily involves
some interaction between the system and an apparatus; hence the usual formalism must be
suitably modified, leading to a joint probability amplitude for the system's time-evolution
and measurement outcomes. The presence of a measuring device can be incorporated by using a
filter function, as developed in the path-integral formulations of continuous quantum
measurement by Mensky~\cite{Mensky1979,Mensky1993}, applied to monitored
oscillators in Ref.~\cite{MenskyOnofrioPresilla1991} and reviewed in
Ref.~\cite{Mensky2002RPI}, and Caves~\cite{Caves1}. In Mensky's
restricted-path-integral approach the measuring device weights each Feynman path by its
compatibility with the recorded outcome, which is precisely the role played by the filter
function below. Beginning with an initial wave
function for the system $\psi_0(x(t_0),t_0)$ at time $t_0$, we make a total of $N$
measurements of $x(t)$ at $t_1,t_2,\dots,t_N$ time steps. Then the joint amplitude for the
$N$ samplings to give a sequence of results $\bar x_1,\dots,\bar x_N$
becomes~\cite{Caves1,Caves2}:
\begin{widetext}
\begin{equation}
\Phi(\bar x_1,\dots,\bar x_N;x,t_N)=\int_{t_0}^{(x,t_f)}\!\!\mathcal{D}[x(t)]
\left(\prod_{q=1}^{N}\Upsilon(\bar x_q-x(t_q))\right)
e^{(i/\hbar)S[x(t)]}\psi_0(x(t_0),t_0).
\label{eq:caves}
\end{equation}
\end{widetext}
where the function $\Upsilon(\bar x_q-x(t_q))$ represents the action of the measuring device
on the system. In qualitative terms, the measurement resolution is captured by the variance
of $|\Upsilon(\bar x_q-x(t_q))|^2$. Therefore, an appropriate function $\Upsilon$ will
modify the phases of each of the interfering paths at measurement times $t_j$, leading to a
joint system--measurement record amplitude $\Phi(\bar x_1,\dots,\bar x_N;x,t_N)$. The extent
to which the phases are perturbed is proportional to the measurement resolution. Therefore,
the path integral formalism, after incorporating the action of the measuring apparatus,
provides information about the joint statistics of a sequence of measurements distributed in
time without worrying about the unitary evolution followed by the ``collapse'' of the wave
function. In our framework, the action of the measurement gives each path a nudge by
altering the phase, which can also be interpreted as a modification to the system action
corresponding to the path. Indeed, {writing $\Upsilon$ in its Fourier representation, $\Upsilon(\bar x_q-x)=\int dk\,\tilde{\Upsilon}(k)\,e^{ik(\bar x_q-x)}$, each Fourier component modifies the action at the measurement times:
\begin{equation}
\prod_{q=1}^{N}\left\{e^{\frac{i}{\hbar}\int_{t_{q-1}}^{t_q}L\,d\tau}\,
e^{ik(\bar x_q-x(t_q))}\right\}
\label{eq:upsilon-exp}
\end{equation}
with the Lagrangian of the system denoted $L$. To summarize, the process of measurement
introduces $\Upsilon$, the conditional amplitude of obtaining $\bar x_q$ when the system was
at $x(t_q)$. The consequence of introducing the measuring device is the change of system
action corresponding to the Feynman paths. The formalism above takes into account the entire
sequence of the measurement process, both the strength and the time of measurement. This
sequence of random variables $\{\bar x_q\}_{q=1}^{N}$ is a surrogate for the trajectory one
might observe. For a sufficiently macroscopic system, the dominant contribution to the
probability amplitude comes from the vicinity of the classical paths, and therefore for
simplicity $x(t_q)$ can be taken to be $x_{cl}(t_q)$. Operationally speaking, the above
formalism gives us the probability amplitude to jointly observe a sequence of random
variables $\bar x_q$, centered around $x_{cl}(t_q)$ with a noise inversely proportional to
the measurement resolution. This framework is complementary to the conventional
picture of quantum mechanics with the system evolving according to two rules---unitary
evolution between measurements and the ``collapse'' of the wave function at each
measurement.

The path integral formalism, after incorporating the measurements, is a conceptual device
to give us the joint statistics of time-distributed measurements performed on a quantum
system. Unlike the conventional picture of unitary evolution and collapse, there is no
similar decomposition of the dynamics. The notion of quantum back-action---the effect of an
instantaneous quantum measurement on the subsequent fate of the trajectory---has already
been incorporated in the perturbations to the phases of Feynman paths as described above.

In the spirit of the correspondence principle and the stationary phase approximation, the
question of the classical--quantum transition is answered by finding the extremal solutions
to the action in Eq.~\eqref{eq:upsilon-exp}. Classically, we have access to the joint
statistics in a straightforward manner: it is possible to unambiguously (up to measurement
resolution) determine the path the system takes without disturbing the system. We show that
this notion of non-disturbance is expressed as the separability of the joint statistics: the
joint amplitude for the measurement outcomes and the system must separate into individual
products for a system operating in the classical regime.

\begin{figure}
    \centering
    \includegraphics[height=6 cm,width=8 cm]{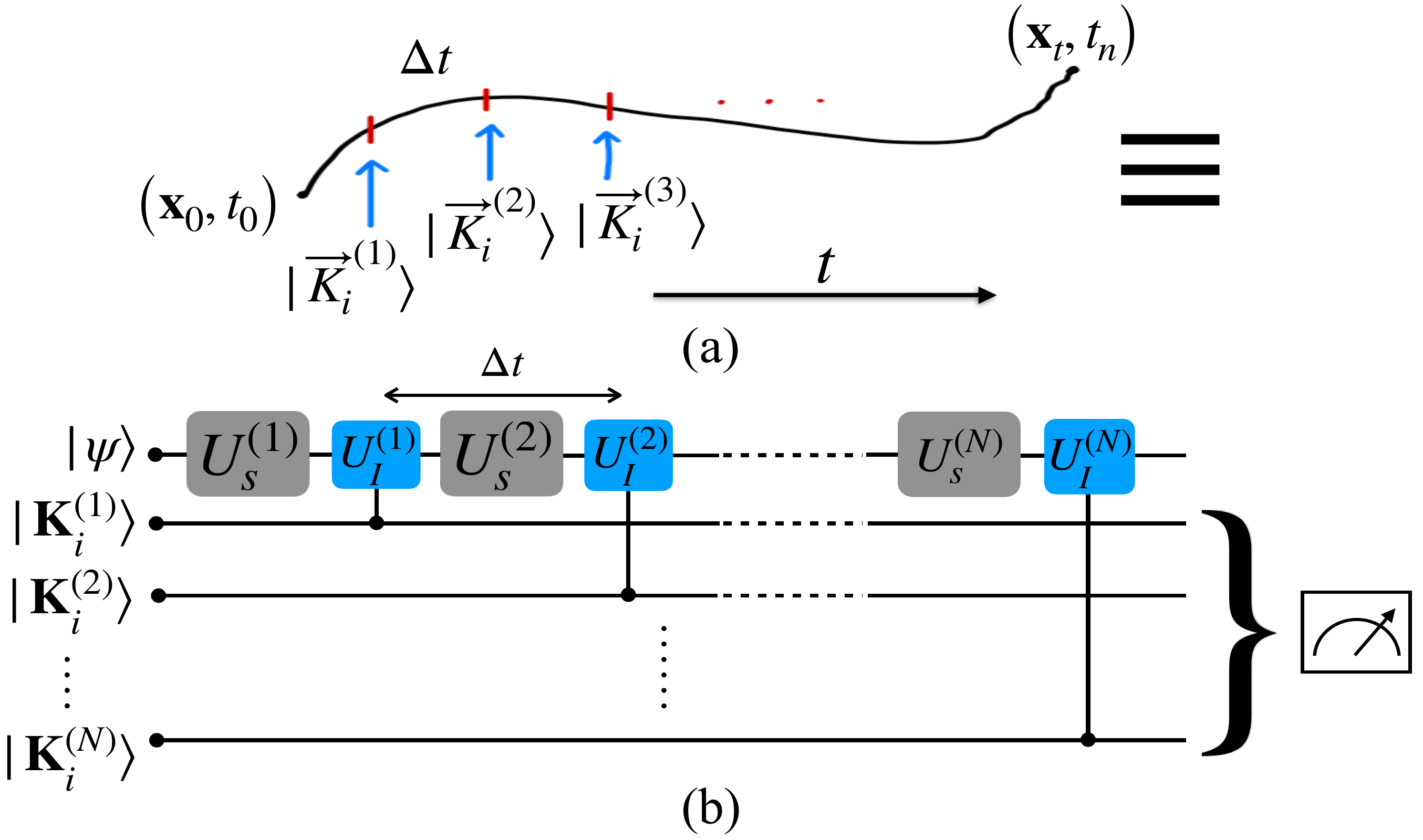}
    \caption{Protocol for measurement distributed in time as a circuit model: (a) At each time step the system is being probed by a photon of a given initial momentum; in between two successive measurements the system is evolving by its free dynamics. We are measuring the scattered photon to get information about the system of interest. This scenario is equivalent to the circuit model shown in (b).}
    \label{measurement_dist_in_time}
\end{figure}

\section{Path integral formalism and the emergence of classicality from standard quantum
mechanical machinery}

One can interpret the aforementioned measurement scheme as being realised physically in the
following way. We start with Eq.~\eqref{eq:caves}, which gives the joint statistics of
multiple measurements performed on the system at different times. We now show how these joint
statistics are calculated by considering a first-principles derivation of the amplitude in
terms of the conventional machinery of quantum mechanics---unitary evolution of the joint
system--probe state. In order to do so, consider the continuous scattering of plane-wave
probes from the system. Consider scattering between a system with momentum $\hbar\mathbf{k}$
and a probe (photon) of momentum $\hbar\mathbf{K}_i$ (here bold symbols imply vectors). This
leads to the photon scattering off to a state with final momentum $\hbar\mathbf{K}_f$ and the
system losing a total momentum $\Delta\mathbf{K}=\mathbf{K}_f-\mathbf{K}_i$. The resolution of
the measurement is inversely proportional to the wavelength of the probe,
$\lambda=2\pi/|\mathbf{K}_i|$. A continuous position record for the system can be obtained by
measuring the probes in the position basis after the scattering. We utilise the formalism for
non-relativistic quantum scattering~\cite{taylor2012scattering}. Consider that a probe, with an initial
momentum $\mathbf{K}_i$, scatters off a system with initial momentum $\mathbf{k}$. In the
momentum basis, the final state is given by:
\begin{equation}
\hat U_{\rm sys-probe}|\mathbf{k}\rangle|\mathbf{K}_i\rangle\to
\int d\mathbf{K}_f\,S(\mathbf{K}_f,\mathbf{K}_i)
|\mathbf{k}-(\mathbf{K}_f-\mathbf{K}_i)\rangle|\mathbf{K}_f\rangle,
\label{eq:U-mom}
\end{equation}
where $S(\mathbf{K}_f,\mathbf{K}_i)$ is the matrix element of the scattering matrix $\hat S$.
Therefore, $S(\mathbf{K}_f,\mathbf{K}_i)$ is the amplitude for the probe with initial momentum
$\mathbf{K}_i$ to be found with momentum $\mathbf{K}_f$ in the final state. The scattering
matrix $\hat S=e^{-i\hat H_I t_I}$, where $\hat H_I$ denotes some interaction between the probe
and the system and $t_I$ denotes the interaction time between the two. By using the scattering
amplitude formalism, we wish to abstract ourselves away from the details of the interaction
Hamiltonian $\hat H_I$. Then, by scattering such probes from the system we expect to recover
the system's position by measuring the probe's position after the
scattering~\cite{TanWalls}. With the system initially in a position eigenstate, the
interaction in Eq.~\eqref{eq:U-mom} leads to the following final state:
\begin{equation}
\hat U_{\rm sys-probe}|\mathbf{r}\rangle|\mathbf{K}_i\rangle\to
\int d\mathbf{K}_f\,S(\mathbf{K}_f,\mathbf{K}_i)
e^{-i(\mathbf{K}_f-\mathbf{K}_i)\cdot\mathbf{r}}|\mathbf{r}\rangle|\mathbf{K}_f\rangle.
\label{eq:U-pos}
\end{equation}
We note that the interaction, beginning with a separable state, leads to a joint state which
is not separable. We aim to make a measurement of the probe in the position basis after the
scattering. Therefore, the joint system--probe wave function, post-scattering, in the position
basis is given as:
\begin{align}
\psi_{\rm tot}(\bar{\mathbf{r}},\mathbf{r})
&=\int d\mathbf{K}_f\,S(\mathbf{K}_f,\mathbf{K}_i)
e^{-i(\mathbf{K}_f-\mathbf{K}_i)\cdot\mathbf{r}}e^{i\mathbf{K}_f\cdot\bar{\mathbf{r}}}
\label{eq:psitot1}\\
&=e^{i\mathbf{K}_i\cdot\bar{\mathbf{r}}}\!\int\! d\mathbf{K}_f\,S(\mathbf{K}_f,\mathbf{K}_i)
e^{-i(\mathbf{K}_f-\mathbf{K}_i)\cdot\mathbf{r}}e^{i(\mathbf{K}_f-\mathbf{K}_i)\cdot\bar{\mathbf{r}}},
\label{eq:psitot2}
\end{align}
which gives the joint probability amplitude for the system to be at position $\mathbf{r}$ and
the probe to be at position $\bar{\mathbf{r}}$ after the interaction. Then, a sequence of
time-distributed measurements on the system can be considered through repeated interactions
with travelling-wave probes. The interaction creates correlations between the two. Measuring
the probes as soon as they interact with the system leads to evolution conditioned on the
outcome, since the correlated state collapses. Instead, we seek the conditional probability
amplitude until the end of the time evolution under the influence of all the probes
simultaneously. We may do this, heuristically, by invoking Eq.~\eqref{eq:caves}, but
considering the Schr\"odinger picture is illuminating. Consider an initially separable joint
state for the system and the probes,
$|\psi\rangle=|\mathbf{r}_i\rangle\prod_{j=1}^{M}\otimes|\phi^{(j)}_{\rm probe}\rangle$. This
state is evolved in time by interrupting the system's evolution with repeated interactions
with the probes (see Fig. \ref{measurement_dist_in_time}):
\begin{widetext}
\begin{equation}
|\psi_f\rangle=\hat U_{\rm sys\text{-}probe}\hat U(t_M,t_{M-1})\hat U_{\rm sys\text{-}probe}
\cdots\hat U_{\rm sys\text{-}probe}\hat U(t_2,t_1)\hat U_{\rm sys\text{-}probe}
\hat U(t_1,t_0)|\mathbf{r}_i\rangle\prod_{j=1}^{M}\otimes|\phi^{(j)}_{\rm probe}\rangle.
\label{eq:joint-evol}
\end{equation}
\end{widetext}
We insert the identity operator $\hat I_j=\int d\mathbf{r}_j|\mathbf{r}_j\rangle\langle
\mathbf{r}_j|$ at times $t_j$ in Eq.~\eqref{eq:joint-evol}, which leads to a path integral
representation in the limits $M\to\infty$ and $\Delta t=t_j-t_{j-1}\to0$, since the
amplitudes $\langle\mathbf{r}_j|\hat U(t_j,t_{j-1})|\mathbf{r}_{j-1}\rangle$ can be expressed
as a path integral. The interaction with the probe only introduces phase factors:
\begin{widetext}
\begin{equation}
\int d\mathbf{K}_f^{(1)}S(\mathbf{K}_f^{(1)},\mathbf{K}_i)\cdots
d\mathbf{K}_f^{(M+1)}S(\mathbf{K}_f^{(M+1)},\mathbf{K}_i)\int d\mathbf{r}_1\cdots d\mathbf{r}_M
\prod_{j=1}^{M+1}\langle\mathbf{r}_j|\hat U(t_j,t_{j-1})|\mathbf{r}_{j-1}\rangle\,
e^{-i(\mathbf{K}_f^{(j)}-\mathbf{K}_i)\cdot\mathbf{r}_j}|\mathbf{K}_f^{(j)}\rangle.
\label{eq:phasefactors}
\end{equation}
\end{widetext}
The additional phase factor $e^{-i(\mathbf{K}_f^{(j)}-\mathbf{K}_i)\cdot\mathbf{r}_j}$
introduced at every time step $\{t_j\}_{j=1}^{M}$ amounts to a local phase change of the wave
function arriving at the point $\mathbf{r}_j$ from $\mathbf{r}_{j-1}$. This results in a
translation of the system's momentum wave function by the amount
$\Delta\mathbf{K}=\mathbf{K}_f^{(j)}-\mathbf{K}_i$, which is the momentum change for the
probes. Similar to the refraction of wave fronts~\cite{Storey1994}, the
outgoing trajectories from $\mathbf{r}_j$ are shifted due to the impulse imparted by the
probes. To proceed to the continuous time limit, we insert the short-time propagator
\begin{widetext}
\begin{equation}
\langle\mathbf{r}_j|\hat U(t_j,t_{j-1})|\mathbf{r}_{j-1}\rangle=
\left(\frac{m}{2\pi i\hbar\Delta t}\right)^{\!3/2}
\!\!e^{\frac{i}{\hbar}\left[\frac{m}{2}\left(\frac{\mathbf{r}_j-\mathbf{r}_{j-1}}{\Delta t}\right)^2\Delta t-V(\mathbf{r}_j)\Delta t\right]}
\label{eq:shorttime}
\end{equation}
\end{widetext}
and arrive at a path integral representation for the sum over system paths. We denote the
product of exponentials $\prod_{j=1}^{M+1}e^{-i(\mathbf{K}_f^{(j)}-\mathbf{K}_i)\cdot\mathbf{r}_j}$ ($M$ scattering events involve $M{+}1$ interaction vertices; for $M\gg1$ we do not distinguish $M$ from $M{+}1$)
as $\exp\!\big(-i\int_0^t\Delta\mathbf{K}(\tau)\cdot\mathbf{r}(\tau) d\tau\big)$ in the continuous limit:
\begin{widetext}
\begin{equation}
Z(\mathbf{r}_f,t_f,\mathbf{r}_i,0;\{\Delta\mathbf{K}\})=
\int_{(\mathbf{r}_i,0)}^{(\mathbf{r}_f,t_f)}\!\!\mathcal{D}\mathbf{r}(\tau)\,
\exp\!\left(\frac{iS[\mathbf{r}(\tau)]}{\hbar}-i\!\int_0^t\!\Delta\mathbf{K}(\tau)\cdot\mathbf{r}(\tau) d\tau \right),
\label{eq:Z-functional}
\end{equation}
\end{widetext}
where $Z(\mathbf{r}_f,t_f,\mathbf{r}_i,0;\{\Delta\mathbf{K}\})$ is the conditional amplitude
for the system to propagate from $(\mathbf{r}_i,t_i=0)$ to $(\mathbf{r}_f,t_f)$ under a
momentum transfer $\hbar\Delta\mathbf{K}(\tau_j)$, at every time step $\tau_j$ from the
probes. Here and below, continuum expressions are shorthand for discrete sums
over scattering events: $\Delta\mathbf{K}(\tau)\equiv\sum_{j}\Delta\mathbf{K}_j\,
\delta(\tau-\tau_j)$, with $[\Delta\mathbf{K}(\tau)]=L^{-1}T^{-1}$, so that
$\int_0^t d\tau\,\Delta\mathbf{K}(\tau)\cdot\mathbf{r}(\tau)
=\sum_j\Delta\mathbf{K}_j\cdot\mathbf{r}(\tau_j)$ is dimensionless.
Measuring the probe's final momentum, as compared to its position, gives
complementary information, that is, which-phase as opposed to which-path information. It is
straightforward to evaluate the functional $Z$ in analytical form for linear systems. For
non-linear systems, semiclassical approximations~\cite{Heller2018} are employed to provide
analytical expressions. Corrections to these expressions can be derived as a series in
$\hbar$~\cite{dewittmorette}.

We proceed by an expansion of the action in Eq.~\eqref{eq:Z-functional} around the
unperturbed classical trajectory $\mathbf{r}_{cl}$ of the system between the points
$(\mathbf{r}_i,0)$ and $(\mathbf{r}_f,t_f)$. Transforming the variables
$\mathbf{r}(\tau)=\mathbf{r}_{cl}(\tau)+\eta(\tau)$, and subsequently expanding the action to
second order, results in:
\begin{widetext}
\begin{equation}
Z(\mathbf{r}_f,t_f,\mathbf{r}_i,0;\{\Delta\mathbf{K}\})=
e^{\frac{iS[\mathbf{r}_{cl}]}{\hbar}}\,e^{-i\int_0^t\Delta\mathbf{K}(\tau)\cdot\mathbf{r}_{cl}(\tau) d\tau}
\int_{(0,0)}^{(0,t_f)}\!\!\mathcal{D}\eta(\tau)\,
\exp\!\left(\frac{i}{2\hbar}\delta^2S[\eta(\tau)]-i\!\int_0^t\!\Delta\mathbf{K}(\tau)\cdot\eta(\tau) d\tau \right).
\label{eq:Z-expand}
\end{equation}
\end{widetext}
{The linear term in $\eta$ shifts the saddle point from $\mathbf{r}_{cl}$ to a recoil-driven
trajectory satisfying $m\ddot{\mathbf{r}}+\nabla V=-\hbar\,\Delta\mathbf{K}(t)$, with the impulsive
momentum transfers $-\hbar\,\Delta\mathbf{K}(t)$ supplied by the probes. In the derivation that
follows we work in the regime in which the accumulated recoil displacement stays below the
coarse-graining length (or wave-packet width), so that the unperturbed classical trajectory
remains the relevant saddle; the quantitative bound on this approximation is given in
Sec.~\ref{sec:window}.}
The variable $\eta(\tau)$ represents the deviation of the system from the unperturbed
classical trajectory $\mathbf{r}_{cl}(\tau)$. Since the probes continuously impart momenta to
the system, this deviation may be significant, depending upon the ratio of the system de
Broglie wavelength (set by the system action) to the wavelength of the probes. Two natural
questions to ask are: first, do classical, unperturbed trajectories survive, even in the case
of continuous measurement? Or does the measurement process completely deviate the system from
its original classical trajectory, resulting in measurement-noise-driven modified
trajectories? Second, how does the information about the trajectory proliferate in the
environment---in other words, what is the mechanism for Darwinism? The ideal regime to work in
is when the measurement does not impart momenta large enough to completely disturb the system
but still provides sufficient trajectory information.

Intuitively, quantum evolution can be visualized as the motion of the system wave function
with its centroid moving along the classical trajectory. However, the wave function will have
a width which fluctuates in time. Then, these deviations in the path integral picture are
simply proportional to the width of the system's wave packet around the classical path. The
greater the spatial extent of the wave function about the classical path, the larger the
deviations. Therefore, in the path integral formalism, the fluctuations of the wave-packet
width are simply captured by adding the contribution from paths nearby the classical path, as
is the case with the integral over $\eta$ in Eq.~\eqref{eq:Z-expand}. Since $\delta^2S[\eta]$
defines a quadratic potential, the width of the wave packet around the classical path is
simply equal to the strength of the zero-point fluctuations of a system in a quadratic
potential:
\begin{equation}
w(\tau)\sim\left(\frac{\hbar}{2\sqrt{m|\nabla^2V(\mathbf{r}_{cl}(\tau))|}}\right)^{\!1/2}.
\label{eq:sigma}
\end{equation}
This heuristic formula shows that the system's wave packet stretches and shrinks depending on
the potential seen by the system at the particular time instant $\tau$. If the wave packet
remains sufficiently localized around the classical path, which is indeed true for a
sufficiently macroscopic system, higher-order corrections to the action $\delta^nS[\eta]$, for
$n\ge3$, can be ignored. Since $\delta^nS\propto\nabla^nV(\mathbf{r}_{cl})$, we demand that the
third (and higher) order change in the potential as the system fluctuates around the classical
paths should be small compared to the second-order change in the
potential~\cite{TanmoyPRL}:
\begin{widetext}
\begin{align}
\frac{|\nabla^nV(\mathbf{r}_{cl}(\tau))|w^n(\tau)} 
{|\nabla^2V(\mathbf{r}_{cl}(\tau))|w^2(\tau)}
&=\frac{|\nabla^nV(\mathbf{r}_{cl}(\tau))|}{|\nabla^2V(\mathbf{r}_{cl}(\tau))|}
\left(\frac{\hbar}{\sqrt{m|\nabla^2V(\mathbf{r}_{cl}(\tau))|}}\right)^{\!\frac{n}{2}-1}
\nonumber\\
&\ll1
\label{eq:higherorder}
\end{align}
\end{widetext}
for $n\ge3$. Starting from an initially localized wave packet, and for quadratic potentials
with $\nabla^3V(\mathbf{r}_{cl})=0$, the above inequality is satisfied at all times. In the
general case of an arbitrary potential, the system remains localized around the classical
trajectory if the potential $V(\mathbf{r})$ does not vary over the width of the wave packet. If
the probe wavelength $\lambda$ is such that $\lambda\gg w_{\max}$, where
$w_{\max}\equiv\max_\tau w(\tau)$ is the largest excursion of the wave-packet width
Eq.~\eqref{eq:sigma} along the trajectory,
a joint measurement of
the system positions can be performed without large back-action and localized trajectory
information can be obtained. 
Two remarks on Eq.~\eqref{eq:sigma} are in order, as it is used repeatedly below. First,
$w(\tau)$ is the zero-point width of the \emph{local} quadratic well $\delta^2S[\eta]$ set by
the curvature $\nabla^2V(\mathbf{r}_{cl}(\tau))$; it is therefore a property of the confinement,
not a fixed number, and coincides with the trap zero-point extent $x_{\rm zpf}$ for a harmonic
potential. Second, for a free or inertial coordinate ($\nabla^2V\to0$) the curvature width
Eq.~\eqref{eq:sigma} diverges: a free packet has no confinement scale and instead \emph{disperses},
its width growing in time rather than settling to a stationary value. In that case the relevant
scale is not a fixed width but the spreading time $t_{\rm sp}\sim m\Delta x^2/\hbar$ over which a
cell-sized packet disperses, and $w_{\max}$ in the condition $\lambda\gg w_{\max}$ is to be
read as the packet width attained over the observation window. This distinction is the origin of
the different form taken by the objectivity ceiling in the two geometries in
Sec.~\ref{sec:window}: for the trap it is set by the fixed ratio $\Delta x/x_{\rm zpf}$, whereas for
the free particle it is set by the time ratio $t_{\rm sp}/\tau_D$.

Therefore, we can simplify $Z(\mathbf{r}_f,t_f,\mathbf{r}_i,0)$ to:
\begin{equation}
e^{\frac{iS[\mathbf{r}_{cl}]}{\hbar}}\,e^{-i\int_0^t\Delta\mathbf{K}(\tau)\cdot\mathbf{r}_{cl}(\tau) d\tau}
\int\mathcal{D}\eta(\tau)\,\exp\!\left(\frac{i}{2\hbar}\delta^2S[\eta(\tau)]\right).
\label{eq:Z-simplified}
\end{equation}
The above expression is valid as long as the resolution of measurement is such that the
probes ignore any deviation of the system from the classical trajectory. Intuitively, this
requires a measurement regime where the probe wavelength is greater than the de Broglie
wavelength of the system. The integral over the fluctuations $\eta(\tau)$ can be performed
exactly by an application of the Gel'fand--Yaglom theorem, leading to:
\begin{equation}
Z(\mathbf{r}_f,t,\mathbf{r}_i,0)=\sqrt{\frac{\partial^2S_{cl}}{\partial r_b\,\partial r_a}}\,
e^{\frac{iS_{cl}}{\hbar}}\,e^{-i\int_0^t\Delta\mathbf{K}(\tau)\cdot\mathbf{r}_{cl}(\tau) d\tau}.
\label{eq:Z-GY}
\end{equation}
Considering a simultaneous measurement of the probes in the position basis, the joint
probability amplitude for the system's propagator and the measurement record
$\{\mathbf{r}_p(\tau),0\le\tau\le t\}$ is obtained as:
\begin{widetext}
\begin{equation}
\Phi(\mathbf{r}_p(t);\mathbf{r}_f,t_f,\mathbf{r}_i,0)=\lim_{M\to\infty}
\exp\!\Big(i\sum_{j=1}^{M}\mathbf{K}_i\cdot\mathbf{r}_p(\tau_j)\Big)
\int\prod_{j=1}^{M+1}d\mathbf{K}_f^{(j)}S(\mathbf{K}_f^{(j)},\mathbf{K}_i)
\exp\!\left(i\!\int_0^t\!\Delta\mathbf{K}(\tau)\cdot\mathbf{r}_p(\tau) d\tau \right)
Z(\mathbf{r}_f,t_f,\mathbf{r}_i,0),
\label{eq:Phi-joint}
\end{equation}
\end{widetext}
where the first factor is a pure phase depending on the measurement
record alone; it cancels identically in the record probability
$|\Phi|^{2}$ and is dropped in what follows.


To evaluate Eq.~\eqref{eq:Phi-joint} we must specify the scattering matrix. We assume that
the system--probe interaction is weak, so that the scattering is elastic and the final
momenta lie on a sphere of radius $|\mathbf{K}_i|$. 
Taking the probes to scatter isotropically, a convenient elastic scattering kernel for the
isotropic model considered here is
\begin{equation}
{S(\mathbf{K}_f,\mathbf{K}_i)=\frac{\delta(|\mathbf{K}_f|-|\mathbf{K}_i|)}{|\mathbf{K}_i|^2}\,A(\hat{\mathbf{K}}_f,\hat{\mathbf{K}}_i)}
\label{eq:Sansatz}
\end{equation}
with the angular amplitude normalized as $\int d\Omega\,|A|^2=1$, and the isotropic choice
$A=1/\sqrt{4\pi}$ corresponding to a uniform distribution of outgoing directions on the unit
sphere.
 Carrying out the integral over the probe directions at each
time step $\tau_j$ (see Appendix~\ref{app:kernel}) yields a sinc kernel,
\begin{equation}
I_j={4\pi}\,e^{-i\frac{2\pi}{\lambda}\delta r_3(\tau_j)}\,
\frac{\sin\!\big(\frac{2\pi}{\lambda}|\mathbf{r}_p(\tau_j)-\mathbf{r}(\tau_j)|\big)}
{\frac{2\pi}{\lambda}|\mathbf{r}_p(\tau_j)-\mathbf{r}(\tau_j)|},
\label{eq:sinc}
\end{equation}
{the angular integral over the full sphere being real, with the incident-momentum phase
$e^{-i\mathbf{K}_i\cdot(\mathbf{r}_p-\mathbf{r})}$ kept separate.} This kernel is 
strikingly similar to the distribution of light diffracted through a pinhole. The series of
probes at times $\tau_j$ essentially scatters off the system, which is at position
$\mathbf{r}_{cl}(\tau_j)$. To take the continuous limit ($M\to\infty$, $\Delta t\to0$) we
scale the resolution as $\lambda=\alpha/\sqrt{\Delta t}$, so that while the number of
measurements in the short time $\Delta t$ blows up, each measurement becomes coarser
($\lambda\to\infty$) in such a manner that the product $\lambda\sqrt{\Delta t}=\alpha$ converges. The
product of the sinc factors over all time steps then collapses---only the leading term
survives, the rest vanishing as $\Delta t\to0$ (see Appendix~\ref{app:kernel})---to a
Gaussian suppression of records that stray from the classical path. Combining with the
semiclassical propagator Eq.~\eqref{eq:Z-GY} gives (without the
additional phase term):
\begin{widetext}
\begin{equation}
\Phi(\mathbf{r}_p;\mathbf{r}_f,t_f,\mathbf{r}_i,0)\propto
\sqrt{\frac{\partial^2S_{cl}}{\partial r_b\,\partial r_a}}\,e^{\frac{iS_{cl}}{\hbar}}
\exp\!\left(-\frac{2\pi^2}{3\alpha^2}\!\int d\tau\,|\mathbf{r}_p(\tau)-\mathbf{r}_{cl}(\tau)|^2\right).
\label{eq:central}
\end{equation}
\end{widetext}
Equation~\eqref{eq:central} is the central result of our work. 

The Gaussian factor is the restricted-path-integral weight functional of Mensky ~\cite{Mensky1979,Mensky1993},
which was used with a phenomenological strength $\kappa$ for continuous position monitoring
~\cite{MenskyOnofrioPresilla1991,Mensky2002RPI}}. Equation~\eqref{eq:central} is its stationary-phase evaluation with
$\kappa=2\pi^{2}/3\alpha^{2}$. Weights of this form have also been derived from sequences of weak
instantaneous meter couplings ~\cite{Mensky2000book,PresillaOnofrioTambini,Caves1,Caves2}, and are expected to arise generally for a
large number of weak, short measurements ~\cite{Mensky2002RPI}. In our derivation the weight originates from
a scattering environment. The strength is fixed by the probe wavelength and the scattering rate,
and the readout is given by the measured momenta of the scattered probes, so that its statistics
are derived and not assumed. For what follows it is important that the scattering which produces
the record also decoheres superpositions, distributes the record over many probes that can be read
separately, and imparts a recoil to the system. These features are absent in the restricted path
integral, which is model independent by construction, and they allow us to discuss the objectivity
of a trajectory and its cost.

\begin{figure}[]
    \centering
    \includegraphics[scale = 0.2]{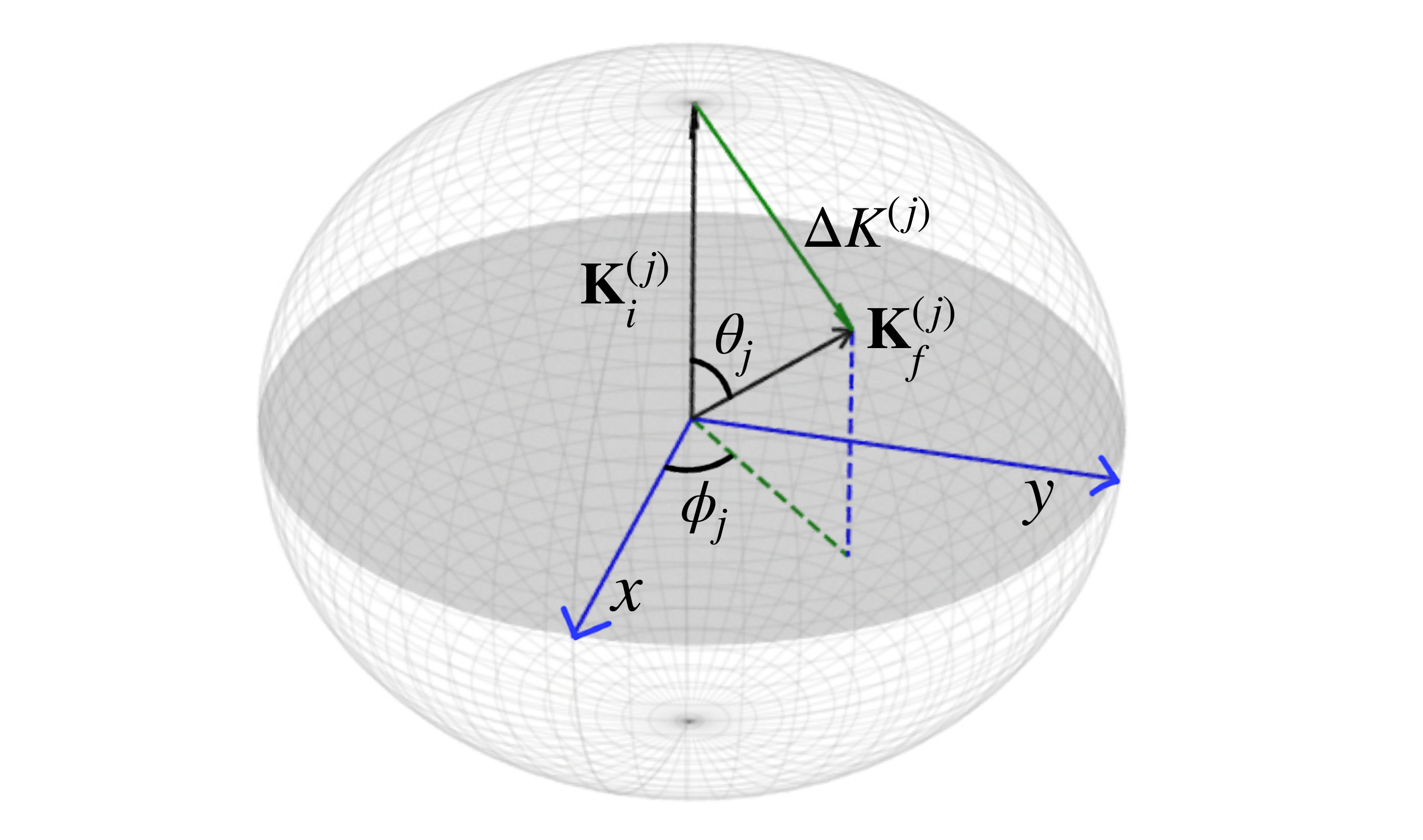}
    \caption{At the $j^{th}$ time step, the system-probe scattering distributes the probe final momentum $\mathbf{K}_{f}^{(j)}$ equally in all directions, leading to the proliferation of information throughout the environment.}
    \label{scatter}
\end{figure}

The prefactor
$\sqrt{\partial^2 S_{cl}/\partial r_b\partial r_a}\,e^{iS_{cl}/\hbar}$ is the semiclassical
amplitude of the system propagator.
The trajectories in the
vicinity of the classical path are robust despite the inevitable disturbance caused by
measurements. While the phases of each of the trajectories are perturbed and averaged over
the probe final momentum, the classical path remains the fixed point of this phase
scrambling. Therefore, repeated interactions with the probes proliferate information about
the classical path throughout the environment, Fig.~\ref{scatter}. This suggests
another way to interpret quantum Darwinism~\cite{Zurek2009,e24111520,PhysRevA.73.062310}. 

We now clarify the role of the measurement in this regime. For a macroscopic system undergoing
regular motion, the free spreading time of the wave packet is very long, so an initially localized
packet does not disperse over any relevant observation time. The dominant contribution to
Eq.~\eqref{eq:central} then comes from the classical path and its immediate neighbors, which still
interfere coherently within the packet width. This holds only for regular motion. For a chaotic
macroscopic coordinate, an isolated packet is stretched exponentially and spreads over macroscopic
scales within an Ehrenfest time ~\cite{BermanZaslavsky1978,ZurekPaz1994}. This time is of the order of seconds to minutes
for everyday chaotic systems, and in that case it is the continuous record that keeps the system
localized, as in Ref.~\cite{TanmoyPRL}. In both cases the measurement provides a record of the path
which is redundant and available to many observers. This record has a cost. If the measurement
were a perfect readout of a pre-existing trajectory, with no disturbance, the redundancy could
grow without bound. The finite back-action ceiling derived in Sec.~\ref{sec:window} shows that a
which-path record always comes with a recoil. Thus, in the semiclassical regime the classical
trajectory is already selected by the dynamics, while its objectivity, i.e.\ the agreement of many
observers on it, is produced by the recording process and is limited by the recoil.

 This identification of the decohering and broadcasting kernels is
specific to \emph{pure} probe environments. For thermal environments the
two separate: the environment decoheres at a finer lengthscale than
information can be retrieved from it, and part of the deposited record
remains bound in thermal noise~\cite{LeeKorbicz2024}. Our conclusions
therefore apply to the pure-probe setting realized by scattering
environments; whether trajectory-level redundancy survives this thermal
readability gap is an open question.

By relating the measurement resolution parameter $\alpha$ to the physical parameters
governing collisional decoherence, we can directly connect our formalism to the notion of
environmental redundancy. In our framework, the total number of independent environmental
probes is naturally identified as
\begin{equation}
M=\frac{t}{\Delta t},
\label{eq:Mprobes}
\end{equation}
where $\Delta t$ denotes the effective interval between successive system--environment
interactions. The continuous-measurement limit of our model is defined by the scaling
$\lambda=\alpha/\sqrt{\Delta t}$, which ensures that the cumulative decoherence exponent
remains finite as $M\to\infty$. To establish a correspondence with physical decoherence, we
equate the exponential suppression factor appearing in Eq.~\eqref{eq:central} with the
standard decoherence factor $\Gamma(t)=\exp(-t/\tau_D)$. For collisional decoherence, the
decoherence rate associated with spatial superpositions of separation $\delta r$ is given by
\begin{equation}
\frac{1}{\tau_D}=\frac{2\pi^2}{3}\frac{\delta r^2}{\alpha^2}.
\label{eq:tauD-rate}
\end{equation}
Substituting $\alpha^2=\lambda^2\Delta t$ yields
\begin{equation}
\tau_D=\frac{3\lambda^2\Delta t}{2\pi^2\delta r^2},
\label{eq:tauD}
\end{equation}
demonstrating that the decoherence timescale scales linearly with the effective interaction
interval $\Delta t$, up to geometric and resolution-dependent factors. The decohering object is the per-event environmental overlap $e^{-\delta r^{2}/4\sigma^{2}}$,
so that $\tau_D=4\sigma^{2}\Delta t/\delta r^{2}$, consistent with Eq.~\eqref{eq:tauD}.
Riedel and Zurek estimate that a micron-sized dust
grain illuminated by sunlight has its position recorded $\sim10^8$ times in
$1\,\mu$s~\cite{PhysRevLett.105.020404}; through their Eq.~(15), and with $\tau_D$ defined here
on the amplitude overlap (twice theirs), this corresponds to $\tau_D\approx10^{-14}\,$s.\footnote{The estimate of Ref.~\cite{PhysRevLett.105.020404} uses dipole (Rayleigh)
scattering, which assumes a grain much smaller than the wavelength. For a micron-sized
grain in sunlight the scattering rate is instead $\Gamma\approx\Phi\,Q_{\rm sca}\,\pi a^2$,
with solar photon flux $\Phi\approx6\times10^{21}\,$m$^{-2}$s$^{-1}$ and Mie efficiency
$Q_{\rm sca}\approx2$--$3$~\cite{BohrenHuffman1983}, i.e.\ $\Gamma\sim10^{10}\,$s$^{-1}$:
$\sim10^4$ complete records per $\mu$s, with $\sim10^8$ reached after a few ms.} The total number of imprinted
environmental records accumulated over an observation time $t$ therefore scales as
\begin{equation}
M=\frac{3}{2\pi^2}\,\frac{t}{\tau_D}\left(\frac{\lambda}{\delta r}\right)^{\!2}.
\label{eq:Mscaling}
\end{equation}
For an observation time $t=1\,\mu$s and spatial resolution $\delta r$ comparable to the probe
wavelength $\lambda$, this yields $M\sim10^8$ independent environmental records.
This is a consistency check on timescales rather than an independent estimate. With
$\delta r\sim\lambda$ every scattered photon is a complete which-path record, so
$\tau_D\approx\Delta t$, the count reduces to the ratio $t/\tau_D$, and the redundancy is
$R\approx M$; it is this $R$, not $M$, that compares with the $\sim10^8$ of Riedel and
Zurek~\cite{PhysRevLett.105.020404}. The $\tau_D\sim10^{-14}\,$s used here is itself obtained from
their estimate~\cite{PhysRevLett.105.020404}. The agreement therefore
confirms that the path-integral record is being counted on the same clock as the photon-scattering
estimate, and should not be read as an independent derivation of the spatial redundancy.

\section{Quantum Darwinism and the redundant encoding of classical trajectories}
\label{sec:gaussianQD}

The phase scrambling mechanism derived above demonstrates how repeated weak scattering events
imprint information about the classical trajectory $\mathbf{r}_{cl}(t)$ onto environmental
probes. In the framework of quantum Darwinism, objectivity is quantified operationally through
the redundancy of classical information, as measured by the quantum mutual information between
the system and fragments of the environment. Given the joint system--environment density matrix
$\rho_{SE}$, the quantum mutual information between the system $S$ and an environmental fragment
$F$ is
\begin{equation}
I(S:F)=H(\rho_S)+H(\rho_F)-H(\rho_{SF}),
\label{eq:MI-def}
\end{equation}
where $H(\rho)=-\mathrm{Tr}(\rho\log\rho)$ denotes the von Neumann entropy. The environment
consists of $M=t/\Delta t$ probes, which we partition into disjoint fragments $F_f$, each
containing a fraction $f$ of the total. Starting from the semiclassical result
Eq.~\eqref{eq:central}, we now make this operational picture explicit, deriving (i) the
classical mutual information from the Gaussian functional, (ii) the emergence of the pointer
basis, and (iii) the Darwinian redundancy plateau recovered once the records are coarse-grained
to bounded classical alternatives, together with the closed-form relation it implies between
redundancy and the decoherence rate, and (iv) a back-action bound on how objective an individual
trajectory can be.

\subsubsection{Classical Mutual Information from the Gaussian Functional}
\label{sec:classMI}

Ignoring the overall phase in Eq.~\eqref{eq:central}, the probability functional for a
measurement record $\mathbf{r}_p(\tau)$ conditioned on a classical trajectory
$\mathbf{r}_{cl}(\tau)$ is
\begin{equation}
P[\mathbf{r}_p|\mathbf{r}_{cl}]\propto\exp\!\left(-\frac{4\pi^2}{3\alpha^2}\!\int_0^t\!
d\tau\,|\mathbf{r}_p(\tau)-\mathbf{r}_{cl}(\tau)|^2\right).
\label{eq:Pfunctional}
\end{equation}

This corresponds to a Gaussian measurement channel.
The continuous functional in
Eq.~\eqref{eq:Pfunctional} defines a noise \emph{intensity} (here $[\,\cdot\,]$ denotes physical dimensions, with $L$ length and $T$ time)
\begin{equation}
\sigma^2_{\rm cont}=\frac{3\alpha^2}{8\pi^2},\qquad[\sigma^2_{\rm cont}]=L^{2}T,
\label{eq:sigma2}
\end{equation}
which, through the continuum scaling $\lambda=\alpha/\sqrt{\Delta t}$, fixes the per-probe
spatial variance
\begin{equation}
{\sigma^2\equiv\sigma^2_{\Delta t}=\frac{\sigma^2_{\rm cont}}{\Delta t}=\frac{3\lambda^2}{8\pi^2},\qquad[\sigma^2_{\Delta t}]=L^{2}.}
\label{eq:sigma2dt}
\end{equation}

The monochromatic probe is a generalized scattering state, and $|\mathrm{sinc}(Kr)|^{2}$ is not
normalizable and has no finite second moment. Equation~\eqref{eq:sigma2dt} should therefore be understood
as the width of the central lobe and not as a variance. We model the per-probe record as a
normalized Gaussian measurement of position with variance $\sigma^{2}$ given by this width. A
finite probe bandwidth regularizes the tails and does not change this width. The scattered probe
state itself is normalized, with overlap $\mathrm{sinc}(K|\mathbf r-\mathbf r'|)$, and no
regularization is needed for the decoherence calculation. In what follows the record statistics
are computed with $\sigma^{2}$ and the decoherence with the full overlap.

Discretising time into $M=t/\Delta t$ independent scattering events, each probe yields
\begin{equation}
r_p^{(j)}=r_{cl}^{(j)}+\xi_j,
\label{eq:channel}
\end{equation}
 where $\xi_j\sim\mathcal{N}(0,{\sigma^2_{\Delta t}})$; here and below
$\sigma^2$ denotes the per-probe variance $\sigma^2_{\Delta t}$.

This is a classical Gaussian channel. 
Let the classical trajectory distribution have variance $\Sigma_S$. This is the prior uncertainty
in the initial conditions of the trajectory, i.e.\ the spread of positions that the observer has
to distinguish, and it can be taken to be the spatial width of the initial wave packet. It is a
property of the state and does not depend on the measurement noise $\sigma^{2}$ or on the
coarse-graining scale $\Delta x$ introduced below. The mutual information from a single probe is
then \cite{6773024}

\begin{equation}
I_1=\frac{1}{2}\log\!\left(1+\frac{\Sigma_S}{\sigma^2}\right).
\label{eq:I1-shannon}
\end{equation}
Throughout the information-theoretic formulas $\log$ denotes $\log_2$, so information is measured in bits.

Here the signal is a one-dimensional label common to all probes, for example a rigid offset of the
candidate trajectory drawn with prior variance $\Sigma_S$. For such a label the Gaussian-channel
results below are exact. For a history with several parameters $\bm\theta=(x_0,v_0,\dots)$ they
generalize to $\tfrac12\log\det\!\big[\mathbb{I}+\Sigma_\theta H^{T}R^{-1}H\big]$, where $H$ is the trajectory design
matrix and  $\Sigma_\theta$ the prior covariance of the parameters and $R$ the noise covariance of the record.

Crucially, all $M$ probes interrogate the same classical position, so they share a common
signal and are not $M$ independent \emph{sources of information} about the system; their
measurement noises $\xi_j$, however, are independent. The sufficient statistic for the record is
therefore the sample mean, whose noise variance is reduced to $\sigma^2/M$, while the single
underlying position keeps the recovered information logarithmic rather than linear in $M$.
 The total mutual information is therefore
\begin{equation}
I(S:E_M)=\frac{1}{2}\log\!\left(1+\frac{M\,\Sigma_S}{\sigma^2}\right),
\label{eq:IM-shannon}
\end{equation}

which grows only logarithmically in $M$ and does not saturate, since a continuous signal has
no finite information ceiling.
 This must be distinguished from the Fisher information, which \emph{is} additive: the
per-probe Fisher information sums to $\mathcal{F}_M=M/\sigma^2$, so the estimation precision and
the signal-to-noise ratio $M\Sigma_S/\sigma^2$ grow linearly, whereas the mutual information is
only their logarithm. Estimation precision (Fisher-linear, unbounded) and redundant objectivity
(set by the bounded record) are therefore distinct notions. For a continuously valued signal the differential entropy
\begin{equation}
H(S)=\frac{1}{2}\log(2\pi e\,\Sigma_S)
\label{eq:HS}
\end{equation}
is \emph{not} an information ceiling: conditional differential entropy is unbounded below, and indeed $I(S:E_M)$ of Eq.~\eqref{eq:IM-shannon} exceeds $H(S)$ at large $M$. A ``fraction of the total information'' is therefore undefined until a resolution scale is introduced; the coarse-graining below supplies exactly that scale.
A fragment $F_f$ containing a fraction $f$ of the probes correspondingly carries
\begin{equation}
I(S:F_f)=\frac{1}{2}\log\!\left(1+\frac{fM\,\Sigma_S}{\sigma^2}\right),
\label{eq:MI-frag2}
\end{equation}
which is concave in $f$ but, being only logarithmic, does not exhibit the sharp ``classical
plateau'' characteristic of bounded or discrete records. The redundancy $R_\delta$ is defined as the number of disjoint fragments that each
supply a fraction $(1-\delta)$ of the total information, $R_\delta=1/f_\delta$, where $f_\delta$
solves $I(S:F_{f_\delta})=(1-\delta)\,I(S:E_M)$. In the high-information regime
$M\Sigma_S/\sigma^2\gg1$, where $\log(1+x)\simeq\log x$, this condition reads
$\log(f_\delta M\Sigma_S/\sigma^2)=(1-\delta)\log(M\Sigma_S/\sigma^2)$, giving
$f_\delta=(M\Sigma_S/\sigma^2)^{-\delta}$ and hence
the redundancy scale
\begin{equation}
R_\delta\sim\left(\frac{M\,\Sigma_S}{\sigma^2}\right)^{\!\delta},
\label{eq:Rscale1}
\end{equation}
a power law in $M$ with exponent set by the information deficit $\delta$, rather than the linear
growth implied by a naive per-probe sum. We emphasize that while the \emph{number} of imprinted
records can be enormous (Eq.~\eqref{eq:Mscaling} gives $M\sim10^8$ for the Riedel--Zurek
estimate), the redundancy of a single continuous Gaussian observable grows only as this weak power,
in contrast to the much larger redundancies obtained from discrete, bounded
records~\cite{PhysRevLett.105.020404}; a quantitative comparison with those estimates therefore
requires care.

\subsubsection{Emergence of the Pointer Basis}

The joint semiclassical state has the approximate branching structure
\begin{equation}
|\Psi(t)\rangle\simeq\sum_\alpha c_\alpha\,{|\psi_\alpha(t)\rangle}
\bigotimes_{j=1}^{M}|E_j(\mathbf{r}_{cl}^{(\alpha)})\rangle.
\label{eq:branch2}
\end{equation}
{Here $|\psi_\alpha(t)\rangle$ is a narrow semiclassical wave packet whose centroid follows the
trajectory $\mathbf{r}_{cl}^{(\alpha)}(\tau)$; the label $\alpha$ is a pointer \emph{history}
rather than an instantaneous position eigenstate. More details on Eq.~\eqref{eq:branch2} are in Appendix~\ref{app:branching}.}

The overlap between environmental states corresponding to two distinct classical trajectories
satisfies
\begin{equation}
\langle E(r)|E(r')\rangle=\exp\!\left(-\frac{M}{4\sigma^2}|r-r'|^2\right).
\label{eq:overlap}
\end{equation}
As $M\to\infty$, this overlap vanishes for $r\ne r'$, causing off-diagonal density matrix
elements to disappear:
\begin{equation}
\rho_S(r,r')\sim\rho_S(r,r')\exp\!\left(-\frac{M}{4\sigma^2}|r-r'|^2\right).
\label{eq:rhodecoh}
\end{equation}
Hence, classical trajectories form the dynamically selected pointer basis.

\subsubsection{Coarse-grained records and the redundancy plateau}
\label{sec:plateau}

The logarithmic growth of Eq.~\eqref{eq:IM-shannon} reflects the fact that a continuous
coordinate can be estimated to ever-finer precision; it is the structure of \emph{parameter
estimation} rather than of objective classical information. Objectivity in the sense of quantum
Darwinism concerns instead a \emph{bounded} alternative: not the value of the trajectory to
arbitrary precision, but \emph{which} of a discrete set of classically distinct, dynamically
selected alternatives---the coarse-grained pointer states identified above---the system
occupies. We therefore coarse-grain the record to the resolution at which positions become
classically distinct.

Consider the simplest such alternative: two \emph{complete candidate histories}
$\mathbf{r}_{\pm}(\tau)=\mathbf{r}_{cl}(\tau)\pm\Delta x/2$, labeled by $s=\pm$---the branching structure of Eq.~\eqref{eq:branch2}
in its simplest form---separated by $\Delta x$ throughout the observation window, with
$H_{\mathcal P}\le\log 2$ the entropy of this history label. More generally, for any two
candidate histories $\alpha,\beta$ the conditional environmental records overlap as
$-\ln|\gamma_{\alpha\beta}|=\frac{1}{4\sigma^{2}}\sum_{j}|\mathbf{r}_{\alpha}(t_j)-\mathbf{r}_{\beta}(t_j)|^{2}
\to\frac{1}{4\sigma^{2}\Delta t}\int dt\,|\mathbf{r}_{\alpha}(t)-\mathbf{r}_{\beta}(t)|^{2}$,
which for the constant-offset pair reduces to the $\gamma_m$ used below. The analogous overlap for thermal oscillator environments, with a
frequency-weighted kernel in place of our memoryless (white) one, was
obtained in Refs.~\cite{TuziemskiKorbicz2015,LeeKorbicz2024}; our
scattering probes realize its Markovian limit.
The redundancy
statements of this section therefore concern a \emph{low-dimensional history label}---the
objectivity question---while the $O(M)$-dimensional fine-grained path remains a matter of
estimation (Sec.~\ref{sec:classMI}). Fragments are taken \emph{time-interleaved}: the $k$-th of $R$
fragments collects every $R$-th scattering event, so that each fragment samples the entire
observation interval and can reconstruct the full history, not one instant of it.

Each probe reports
\begin{equation}
r_p^{(j)}=s\,\frac{\Delta x}{2}+\xi_j,\qquad \xi_j\sim\mathcal{N}(0,\sigma^2),
\label{eq:cg-channel}
\end{equation}
with 
 $\sigma^2={\sigma^2_{\Delta t}=3\lambda^2/8\pi^2}$
 as in Eq.~\eqref{eq:sigma2dt}. A fragment of $m=fM$ probes
identifies $s$ through the sample mean, whose noise variance is $\sigma^2/m$; the optimal
(Bayes) probability of misassigning the pointer is
\begin{equation}
P_e(m)=Q\!\left(\sqrt{m}\,\frac{\Delta x}{2\sigma}\right)\le\frac{1}{2}\,
e^{-m\,\Delta x^2/8\sigma^2},
\label{eq:cg-error}
\end{equation}
with $Q$ the Gaussian tail function; the bound is the Bhattacharyya estimate. The fragment
information is therefore bounded from below (Fano's inequality) and saturates at the pointer entropy,
\begin{equation}
I(S:F_f)\ge H_{\mathcal P}-H_2\!\big(P_e(fM)\big)\;\xrightarrow{\,fM\gg m_*\,}\;H_{\mathcal P},
\label{eq:cg-MI}
\end{equation}
where $H_2$ is the binary entropy and

\begin{equation}
m_*\sim\frac{8\sigma^2}{\Delta x^2}={\frac{3\lambda^2}{\pi^2\,\Delta x^2}}
\label{eq:cg-depth}
\end{equation}

\begin{figure}[t]
\centering
\includegraphics[width=0.96\linewidth]{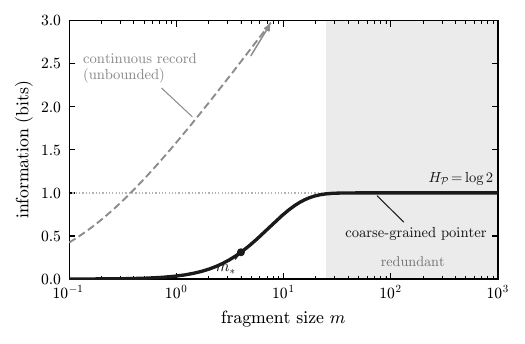}
\caption{Partial information versus fragment size $m$. The continuous Gaussian record carries
a mutual information $\tfrac12\log_2(1+m\,\Sigma_S/\sigma^2)$ that grows without bound, reflecting
ever-improving estimation precision, whereas a coarse-grained binary pointer record saturates at
the pointer entropy $H_{\mathcal P}=\log_2 2$ once the fragment exceeds a few times the information depth
$m_*$, reflecting redundant, objective information. The depth $m_*$
(marked) sets the scale on which the misassignment probability decays exponentially rather than the
saturation point itself. The pointer curve is the Bhattacharyya lower bound of Eq.~\eqref{eq:cg-error}:
it guarantees $0.31\,H_{\mathcal P}$ at $m=m_*$ and completeness to within $1\%$ by $m\simeq6m_*$
(shaded); the exact values are $0.72\,H_{\mathcal P}$ and $m\simeq4m_*$.
Parameters $\Sigma_S/\sigma^2=8$ and
$\Delta x^2/\sigma^2=2$.}
\label{fig:partialinfo}
\end{figure}
is the information depth---the number of probes required to resolve the two cells. In contrast
to the unbounded estimate Eq.~\eqref{eq:IM-shannon}, the information now \emph{saturates}: once a
fragment holds more than a few times $\sim m_*$ probes it already carries essentially the whole classical
record $H_{\mathcal P}$, and additional probes are redundant. More precisely, $m_*$ sets the
scale on which $P_e$ decays exponentially rather than the saturation point itself: at $m=m_*$ the
Bhattacharyya bound gives $P_e\le\tfrac12 e^{-1}$ and $I(S:F_f)\ge0.31\,H_{\mathcal P}$ (exactly,
$0.72\,H_{\mathcal P}$), and $99\%$ of $H_{\mathcal P}$ is reached by $m\simeq4m_*$ ($6m_*$ for the bound).
This is exactly the classical
plateau of the partial-information plot that serves as the operational signature of quantum
Darwinism (Fig. \ref{fig:partialinfo}).

Crucially, $m_*$ is fixed by the per-probe resolution $\Delta x/\sigma$ alone and is therefore
\emph{independent of $M$}. The fragment size $m_\delta$ that supplies $(1-\delta)H_{\mathcal P}$
is consequently an $M$-independent constant of order $m_*$, and the redundancy scales
\emph{linearly} with the number of records,
\begin{equation}
R_\delta=\frac{M}{m_\delta}\propto M=\frac{t}{\Delta t},
\label{eq:cg-redundancy}
\end{equation}
in sharp contrast to the weak power law $R_\delta\sim(M\Sigma_S/\sigma^2)^{\delta}$ of
Eq.~\eqref{eq:Rscale1}. The boundedness of the record is precisely what converts
diminishing-returns estimation into massively redundant, objective information. Inserting the
$M\sim10^8$ scattering events estimated in Eq.~\eqref{eq:Mscaling}, Eq.~\eqref{eq:cg-redundancy}
gives $R_\delta\sim10^8$, taking the Riedel--Zurek
estimate~\cite{PhysRevLett.105.020404} at face value --- a shared-timescale check, as noted
below Eq.~\eqref{eq:Mscaling}, rather than an independent route to their number.

\subsubsection{A localization--redundancy relation}
\label{sec:locred}

The plateau of Eq.~\eqref{eq:cg-MI} was obtained from the classical discrimination of the
records. We now sharpen it with the accessible (Holevo) information, which exposes a direct relation
between the redundancy and the decoherence rate. The essential observation is that a single
object---the environmental overlap of Eq.~\eqref{eq:overlap}---controls \emph{both} the loss of
coherence and the information carried by a fragment. Writing
$\gamma_m\equiv\langle E_+^{F}|E_-^{F}\rangle=e^{-m\,\Delta x^2/4\sigma^2}$ for the overlap of
the two conditional environmental states restricted to a fragment of $m$ probes, the
decoherence of the pointer after all $M$ probes is governed by $\gamma_M$, while the information a fragment carries about the pointer is the Holevo quantity of the two
(pure) conditional records,
\begin{equation}
I(S:F_m)=H_2\!\left(\frac{1+\gamma_m}{2}\right),
\label{eq:locred-chi}
\end{equation}
which rises monotonically from $0$ to $H_{\mathcal P}=\log 2$ as $\gamma_m$ runs from $1$ to
$0$, reproducing the plateau of Eq. ~\eqref{eq:cg-MI}. This is the accessible \emph{classical} pointer information, bounded by $H_{\mathcal P}$; it is not the full quantum mutual information of the globally pure branching state, which for a fragment with complement overlap $\gamma_{M-m}$ reads $I_Q(S{:}F_m)=H_2\big(\tfrac{1+\gamma_M}{2}\big)+H_2\big(\tfrac{1+\gamma_m}{2}\big)-H_2\big(\tfrac{1+\gamma_{M-m}}{2}\big)$ and approaches $2H_{\mathcal P}$ for the whole environment. In the Darwinian plateau regime ($\gamma_M\to0$ with $\gamma_{M-m}\to0$) the two coincide, which is the regime used here.

Decoherence provides the natural clock: the coherence of the pointer is suppressed as
$\gamma_M=e^{-t/\tau_D}$, so that
\begin{equation}
\frac{t}{\tau_D}=-\ln\gamma_M=M\,(-\ln\gamma_1),
\label{eq:locred-clock}
\end{equation}
with $\tau_D$ the decoherence time of Eq.~\eqref{eq:tauD}. The redundancy follows the standard
definition $R_\delta=M/m_\delta$, where the threshold fragment $m_\delta$ supplies a fraction
$(1-\delta)$ of the record, $I(S:F_{m_\delta})=(1-\delta)H_{\mathcal P}$. By
Eq.~\eqref{eq:locred-chi} this fixes a threshold overlap $\gamma_\delta$ through
$H_2\!\big((1+\gamma_\delta)/2\big)=(1-\delta)\log 2$, whence
$m_\delta=(-\ln\gamma_\delta)/(-\ln\gamma_1)$. Substituting into $R_\delta=M/m_\delta$ and using
Eq.~\eqref{eq:locred-clock}, the per-probe factor $-\ln\gamma_1$, which carries all of the
microphysics ($\Delta x$, $\sigma$, the scattering rate), cancels and we get
\begin{equation}
R_\delta(t)=\frac{1}{-\ln\gamma_\delta}\,\frac{t}{\tau_D}\equiv A(\delta)\,\frac{t}{\tau_D}.
\label{eq:locred}
\end{equation}
The redundancy is thus proportional to the number of elapsed decoherence times, with a
prefactor $A(\delta)=1/(-\ln\gamma_\delta)$ that depends only on the chosen objectivity
threshold and on no parameter of the system--probe coupling. At the standard value
$\delta=0.1$ one finds $\gamma_\delta\simeq 1/e$, so that $A\simeq 1$ and
\begin{equation}
R_{0.1}\simeq\frac{t}{\tau_D};
\label{eq:locred-std}
\end{equation}
one independent record proliferates per $e$-fold of coherence loss (for comparison,
$A(0.01)\simeq0.47$ and $A(0.5)\simeq4.0$). This is the same linear-in-$M$ redundancy of
Eq.~\eqref{eq:cg-redundancy}, now with its proportionality constant fixed in closed form. As a
consistency check, Ref.~\cite{PhysRevLett.105.020404} defines its decoherence factor as
$\Gamma=|\langle E_1|E_2\rangle|^2=e^{-t/\tau_D}$, so our $\tau_D$ is twice theirs; with this,
Eq.~\eqref{eq:locred} reproduces their Eq.~(15),
$A(\delta)=2/\ln[(2\delta\ln2)^{-1}]=1.01$ at $\delta=0.1$. Taking the Riedel--Zurek estimate at
face value, a micron-sized dust grain decoheres on a timescale $\tau_D\sim10^{-14}\,$s,
so that over an observation time $t=1\,\mu$s Eq.~\eqref{eq:locred-std} gives
$R_{0.1}\sim10^{8}$, with the same caveat: $\tau_D$ is taken from their estimate. The linear rise of redundancy with the number of
decoherence times has been established for spin environments~\cite{RiedelZurekZwolak2012} and for
nonideal, noisy channels~\cite{ZwolakQuanZurek2009,ZwolakQuanZurek2010}, and the link between
redundancy and the optimal distinguishability of pointer records is made precise by the quantum
Chernoff information~\cite{ZwolakRiedelZurek2014}; Eq.~\eqref{eq:locred} is the Gaussian
specialization of this structure in the present scattering setting, with a prefactor $A(\delta)$
set by the objectivity threshold.

Two points are worth noting. First, localization and redundancy are not independent, since both follow from the same
overlap. An environment that decoheres the system also broadcasts a proportionate amount
of redundant information, and the ratio in Eq.~\eqref{eq:locred} is fixed by the scattering geometry.
 Within the ideal pure-scattering model considered here, then, decoherence is accompanied by
proportional information broadcasting; mixed, absorptive, or otherwise inaccessible channels can
decohere a system without producing accessible, redundant records.
Second, Eq.~\eqref{eq:locred} assumes the fully decohered regime, in which the conditional
environmental records are effectively pure, and was written for a binary pointer. The construction
extends directly to an alphabet of $\mathcal{N}$ equally spaced cells: the conditional records
acquire pairwise overlaps $\gamma_m^{(a-b)^2}$ and the fragment information saturates at
$H_{\mathcal P}=\log\mathcal{N}$. Carrying the same analysis through replaces $A(\delta)$ by
\begin{equation}
A(\delta,\mathcal{N})=\frac{1}{-\ln\gamma_\delta}
=\frac{2}{\ln(1/\delta)-\ln\ln\mathcal{N}-\ln\!\frac{\mathcal{N}}{\mathcal{N}-1}},
\label{eq:A-N}
\end{equation}
which again retains no microphysics and grows only weakly with the alphabet size: at
$\delta=0.1$ it runs from $A\simeq1.0$ ($\mathcal{N}=2$) through $1.5$ ($\mathcal{N}=10$) to
$2.6$ ($\mathcal{N}=100$), against exact Holevo values $1.0$, $1.4$ and $2.2$. The relation $R_\delta\simeq A(\delta,\mathcal{N})\,t/\tau_D$ therefore
remains an $O(1)$ multiple of $t/\tau_D$ for any physical pointer alphabet, reducing to
Eq.~\eqref{eq:locred-std} for $\mathcal{N}=2$; the underlying small-overlap expansion is controlled for $\delta\ln\mathcal{N}\ll1$, i.e.\ $\mathcal{N}\ll e^{1/\delta}$ (at $\mathcal{N}=100$, $\delta=0.1$, where $\delta\ln\mathcal{N}\simeq0.46$, it overestimates the exact value by about $20\%$).

\subsubsection{A back-action limit on the objectivity of trajectories}
\label{sec:window}

Equation~\eqref{eq:locred} counts the proliferation of records but not their cost: every probe
that carries away which-path information also delivers a recoil, and both follow from the same
kernel. A probe scattering from $\mathbf{K}_i$ to $\mathbf{K}_f$ imparts a momentum kick
$\delta\mathbf{p}=\hbar(\mathbf{K}_i-\mathbf{K}_f)$; for elastic, isotropic scattering
($|\mathbf{K}_f|=|\mathbf{K}_i|\equiv K$ with $\mathbf{K}_f$ uniform over the sphere) the mean is
a deterministic radiation-pressure term, while the fluctuating part has per-component variance
$\langle\delta p^2\rangle=\hbar^2K^2/3$. With scattering rate $\Gamma=1/\Delta t$ this is a
momentum-diffusion coefficient
\begin{equation}
D_p=\frac{\langle\delta p^2\rangle}{2\,\Delta t}=\frac{\hbar^2K^2}{6}\,\Gamma .
\label{eq:Dp}
\end{equation}
The same scattering sets the decoherence rate of Eq.~\eqref{eq:tauD-rate}, which in these
variables reads $\tau_D^{-1}=\tfrac{1}{6}\Gamma K^2\Delta x^2$, so that diffusion and decoherence
are locked together,
\begin{equation}
D_p\,\tau_D\,\Delta x^2=\hbar^2 .
\label{eq:FDT}
\end{equation}
The geometric and rate factors cancel, so heating and decoherence are governed by the same
scattering and related through $\hbar$. This decoherence--diffusion (information--disturbance) relation bounds the
redundancy a trajectory can carry. No dissipative friction or thermal balance enters, so it is not a fluctuation--dissipation theorem; in an anisotropic setting it generalizes to the tensor form $\tau_D^{-1}=\Delta x_i D_{ij}\Delta x_j/\hbar^{2}$. A relation of the same information-gain--disturbance type,
$\lambda_{dB}\lambda_{dist}=\hbar/M\Lambda$, was derived for thermal
environments in Ref.~\cite{LeeKorbicz2024}.

Localization keeps the conditional wave packet narrow at each instant, but the recoil makes its
centroid execute a random walk. For an inertial degree of freedom the accumulated wander of the
centroid due to back-action (denoted by the subscript ``BA'') is
\begin{equation}
\langle\Delta x_{\rm BA}^2(t)\rangle{\;\simeq\;}\frac{2D_p}{3m^2}\,t^3
=\frac{2\hbar^2}{3m^2\tau_D\Delta x^2}\,t^3 ,
\label{eq:BAwander}
\end{equation}
where Eq.~\eqref{eq:FDT} was used, and the derivation of the $t^3$ law is given in
Appendix~\ref{app:backaction}.
The recorded path coincides with the \emph{deterministic}
classical trajectory only while this stays within the resolution,
$\langle\Delta x_{\rm BA}^2\rangle\lesssim\Delta x^2$, that is for
$t\lesssim t_{\rm BA}=\big[3m^2\Delta x^4\tau_D/2\hbar^2\big]^{1/3}$, while objectivity requires
at least one redundant record, $t\gtrsim\tau_D$. An objective, deterministic classical trajectory
therefore exists only within the window
\begin{equation}
\tau_D\;\lesssim\;t\;\lesssim\;t_{\rm BA},
\label{eq:window}
\end{equation}
which is non-empty only if $\tau_D\lesssim t_{\rm sp}\equiv m\Delta x^2/\hbar$, the quantum
spreading time of a cell-sized packet---the localization criterion of Ref.~\cite{TanmoyPRL},
recovered here as the condition for objective trajectories to exist at all. 
We stress that this boundary is the localization condition $\tau_D\sim t_{\rm sp}$, not the
semiclassical condition $S_{cl}\gg\hbar$; for a free coordinate the two are independent. Writing
$S_{cl}\sim\tfrac12 m v^2 t$ along a path of length $L=vt$, one finds
$S_{cl}/\hbar\sim(\Delta x/\lambda_{\rm dB})(L/\Delta x)$ at $t\sim t_{\rm sp}$, which can remain
enormous even as $\tau_D\to t_{\rm sp}$: a macroscopic free particle may sit at the edge of the
objective window while its action is still overwhelmingly classical. For the inertial case,
therefore, the closing of the window marks the loss of \emph{objectivity}, not the onset of the
deep-quantum regime.
The largest
redundancy such a trajectory can carry then follows from Eqs.~\eqref{eq:locred} and
\eqref{eq:window},
\begin{equation}
R_{\max}\simeq A(\delta)\left(\frac{3}{2}\right)^{1/3}\!\left(\frac{t_{\rm sp}}{\tau_D}\right)^{2/3},
\label{eq:Rmax}
\end{equation}
fixed by the kernel. For a micron dust grain ($m\approx10^{-15}\,$kg, $\Delta x\sim1\,\mu$m,
$t_{\rm sp}\sim10^{7}\,$s, $\tau_D\sim10^{-14}\,$s) Eq.~\eqref{eq:FDT} does not apply, since
$K\Delta x\approx13$ and each scattered photon is a complete record ($\tau_D\approx\Delta t$);
with $\langle\Delta x_{\rm BA}^2\rangle=2D_pt^3/3m^2$ and $D_p=\hbar^2K^2/6\Delta t$ this gives
$R_{\max}\simeq t_{\rm BA}/\Delta t\sim4\times10^{13}$ ($\sim10^{11}$ at the Mie scattering
rate), far above any redundancy accumulated in
practice, so back-action is negligible and the trajectory is overwhelmingly objective.
For a typical levitated silica nanosphere (diameter $\approx143\,$nm,
$m\approx3.4\times10^{-18}\,$kg, $\Delta x\sim10\,$nm,
$\tau_D\sim10^{-11}\,$s from the measured recoil heating~\cite{Jain2016} via
Eq.~\eqref{eq:FDT})~\cite{GonzalezBallestero2021} the ceiling remains large,
$R_{\max}\sim5\times10^{7}$. For a macroscopic system the ceiling is therefore far above the redundancy accumulated in
practice, and back-action does not limit the objectivity of the trajectory.
The bound becomes restrictive only as the window itself closes:
$R_{\max}\to\mathcal{O}(1)$ as $\tau_D\to t_{\rm sp}$, which is precisely the localization
criterion of Ref.~\cite{TanmoyPRL} and the edge of the semiclassical regime assumed throughout.
The ceiling should therefore be understood as a consistency condition and not as a practical
limit. Objective, deterministic trajectories exist in the regime where our framework is valid,
and at the boundary of this regime the redundancy reduces to a single record.

The correspondence is quantitative. In the stochastic-measurement description of
Ref.~\cite{TanmoyPRL} the localizing term $-k[\hat X,[\hat X,\rho]]$ suppresses coherence across a
separation $\Delta x$ at a rate $8\eta k\,\Delta x^2$, which is our decoherence rate
$1/\tau_D$ of Eq.~\eqref{eq:tauD-rate} up to the efficiency $\eta$ and factors of order unity.
Their requirement that the band-limited record, averaged over a time $\Delta t$, track the
trajectory to resolution $\Delta x$ [Ref.~\cite{TanmoyPRL}, Eq.~(7): $8\eta k>1/\Delta t\,\Delta
x^2$] is then simply $\tau_D\lesssim\Delta t$, and with the natural choice $\Delta t\sim t_{\rm sp}$
--- averaging just long enough to resolve one cell before it disperses --- it becomes exactly
$\tau_D\lesssim t_{\rm sp}$. For the linear (free and harmonic) dynamics considered here their
curvature condition [Ref.~\cite{TanmoyPRL}, Eq.~(5), $\propto\partial_x^2 F$] is automatically
satisfied, so this record condition is the operative lower edge; their upper, low-noise bound
[Ref.~\cite{TanmoyPRL}, Eq.~(6)] is the back-action limit that we sharpen into the redundancy
ceiling below.

\begin{figure}[t]
\centering
\includegraphics[width=0.96\linewidth]{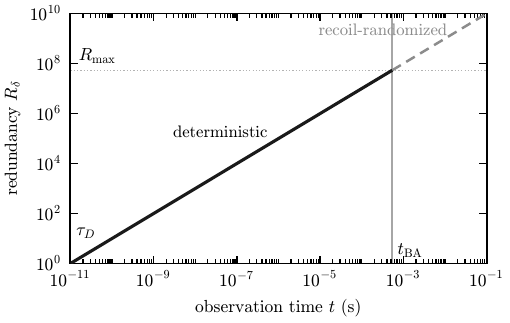}
\caption{{Redundancy of an individual classical trajectory versus observation time, for a
levitated silica nanosphere ($m\approx3.4\times10^{-18}\,$kg, $\Delta x\sim10\,$nm,
$\tau_D\sim10^{-11}\,$s; plotted with $A(\delta)=1$). Objectivity requires at least one redundant record, $t\gtrsim\tau_D$,
while the recorded path stays deterministic only while the recoil-driven wander remains below the
resolution, $t\lesssim t_{\rm BA}\approx0.5\,$ms. Within this window the redundancy grows linearly,
$R_\delta=A(\delta)\,t/\tau_D$, and reaches at most $R_{\max}\approx5\times10^{7}$; the ceiling is simply
the linear law evaluated at $t_{\rm BA}$. Beyond $t_{\rm BA}$ the environment still records the
path redundantly and observers still agree, but the trajectory is recoil-randomized and Brownian
rather than Newtonian. The ceiling falls to a single record only as $\tau_D\to t_{\rm sp}$, that
is, at the edge of the semiclassical regime.}}
\label{fig:redundancy}
\end{figure}

Beyond $t_{\rm BA}$ the environment still records a path redundantly, and observers still agree,
but the path is now recoil-randomized and Brownian rather than deterministic.
Equation~\eqref{eq:window} thus identifies $t_{\rm BA}$ as the point at which an objective
\emph{Newtonian} trajectory gives way to an objective \emph{stochastic} one.
We emphasize that the dynamics underlying this crossover is not itself new: the passage from
ballistic to recoil-diffusive motion driven by photon recoil is standard levitated optomechanics
and has been measured directly~\cite{Jain2016,GonzalezBallestero2021}, and the $D_p$ appearing in
Eq.~\eqref{eq:FDT} is precisely the measured recoil-heating coefficient. What
Eq.~\eqref{eq:window} contributes is an interpretation rather than a new prediction: the same
coefficient that sets the onset of momentum diffusion also fixes the time beyond which the
redundantly recorded path, while remaining objective in the sense of
Sec.~\ref{sec:gaussianQD}, ceases to be deterministic. For the nanosphere above this occurs at
$t_{\rm BA}\approx0.5\,$ms, nearly eight decades after $\tau_D$, with the ceiling
$R_{\max}\sim5\times10^{7}$ nowhere near saturated (Fig.~\ref{fig:redundancy}).
The estimate above assumes an inertial coordinate
(the $t^3$ recoil law) and the heuristic threshold $\langle\Delta x_{\rm BA}^2\rangle\sim\Delta
x^2$, which fix the scaling and the existence of the window rather than the $O(1)$ prefactor.

The experimentally relevant geometry is, however, a trapped one. For harmonic confinement of
frequency $\omega$ the recoil no longer drives a ballistic run-away: heating raises the oscillator
energy linearly, $\dot E=D_p/m$, so the deviation from the deterministic orbit grows as
$\langle\Delta x_{\rm BA}^2(t)\rangle=D_p t/m^2\omega^2$ rather than as $t^3$. The
deterministic-to-Brownian crossover then occurs at $t_{\rm BA}=m^2\omega^2\Delta x^2/D_p$, and the
objectivity ceiling becomes
\begin{equation}
R_{\max}^{\rm trap}\simeq A(\delta)\,(\omega\,t_{\rm sp})^2
=A(\delta)\left(\frac{\Delta x^2}{2x_{\rm zpf}^2}\right)^{\!2},
\label{eq:Rmax-trap}
\end{equation}
with $x_{\rm zpf}=\sqrt{\hbar/2m\omega}$ the zero-point extent. Here the decoherence time cancels: a faster environment supplies records and back-action in the
same proportion, so the trapped ceiling depends only on the resolution measured in units of the
zero-point motion, scaling as $(\Delta x/x_{\rm zpf})^4$ (equivalently, the square of the
recoil-phonon number $n_{\rm BA}=\Delta x^2/2x_{\rm zpf}^2$ at the crossover).
{For the same nanosphere trapped at $\omega/2\pi\approx100\,$kHz ($x_{\rm zpf}\approx5\,$pm)
with $\Delta x\sim1\,$nm this gives $R_{\max}^{\rm trap}\approx4\times10^{8}$. Here too the
ceiling collapses only at the classical--quantum boundary: $R_{\max}^{\rm trap}\to1$ as
$\Delta x\to\sqrt{2}\,x_{\rm zpf}$, that is, when the alternatives the observers are asked to
agree upon cease to be separated by more than the zero-point extent.} The two
geometries therefore give different scalings, $R_{\max}\propto(t_{\rm sp}/\tau_D)^{2/3}$ when
inertial and $R_{\max}\propto(\Delta x/x_{\rm zpf})^4$ when trapped, both following from the same
kernel.
For the trapped geometry the two boundaries do coincide, and this can be made precise. A
harmonic orbit of amplitude $X$ carries an abbreviated action $\oint p\,dx=2\pi E/\omega=\pi m\omega X^{2}$ (the Lagrangian action over a full period vanishes, since $\langle T\rangle=\langle V\rangle$), and using $x_{\rm zpf}^2=\hbar/2m\omega$ this is
\begin{equation}
\frac{S_{cl}}{\hbar}\sim\frac{\pi}{2}\left(\frac{X}{x_{\rm zpf}}\right)^2 .
\label{eq:Scl-zpf}
\end{equation}
Both the objectivity ceiling, $R_{\max}^{\rm trap}\sim(\Delta x^2/2x_{\rm zpf}^2)^2$, and the
action are thus powers of the \emph{same} ratio $\Delta x/x_{\rm zpf}$ (identifying the resolved
scale $X$ with the cell size $\Delta x$). Hence the limit in which the window closes,
$\Delta x\to\sqrt{2}\,x_{\rm zpf}$ and $R_{\max}^{\rm trap}\to1$, is the same limit in which
$S_{cl}/\hbar\to\pi=\mathcal{O}(1)$.

Thus, in the trapped case the redundancy reduces to a single record in the same limit in which
the semiclassical, single-trajectory description breaks down.
 (The numerical factors here are
order-of-magnitude; the identification $X\sim\Delta x$ is the modeling assumption that the
resolved alternatives are separated by about the orbit scale.)

\subsubsection{{A worked example: the free particle}}

{To make the construction concrete, take a free particle ($V=0$), whose classical path is
$\mathbf{r}_{cl}(\tau)=\mathbf{r}_i+\mathbf{v}\,\tau$ with action $S_{cl}=\tfrac12 m v^2 t$.
Equation~\eqref{eq:central} then gives the record likelihood}
\begin{equation}
{P[\mathbf{r}_p|\mathbf{r}_{cl}]\propto\exp\!\left(-\frac{4\pi^2}{3\alpha^2}\!\int_0^t\!d\tau\,
|\mathbf{r}_p(\tau)-\mathbf{r}_i-\mathbf{v}\tau|^2\right),}
\label{eq:free-like}
\end{equation}
{a Gaussian centered on the straight-line trajectory with per-probe variance
$\sigma^2_{\Delta t}=3\lambda^2/8\pi^2$. The continuous record estimates $(\mathbf{r}_i,\mathbf{v})$
with a precision that sharpens as $1/M$ through Eq.~\eqref{eq:IM-shannon}, whereas a coarse-grained
which-cell record saturates at the pointer entropy after a few times $m_*\sim 3\lambda^2/\pi^2\Delta x^2$
probes (Eq.~\eqref{eq:cg-depth}) --- the distinction between estimation precision and redundant
objectivity in its simplest setting. Because $V=0$ the recoil is purely inertial, so the
back-action window of Sec.~\ref{sec:window} applies in its $t^3$ form: the recorded path stays
deterministic for $\tau_D\lesssim t\lesssim t_{\rm BA}=[3m^2\Delta x^4\tau_D/2\hbar^2]^{1/3}$ and
carries at most $R_{\max}\simeq A(\delta)(3/2)^{1/3}(t_{\rm sp}/\tau_D)^{2/3}$ redundant records,
reproducing the {nanosphere ($R_{\max}\sim5\times10^{7}$)}
estimate of Sec.~\ref{sec:window} (the dust grain, with $K\Delta x\gg1$, needs the saturated form
given there).}

The Gaussian suppression derived in
Eq.~\eqref{eq:central} provides a microscopic, trajectory-level mechanism for quantum
Darwinism: repeated weak scattering generates redundant classical records, dynamically
selects pointer trajectories, and drives a crossover from coherent path superposition to
objective classical reality. Operationally, this means that multiple observers can each recover
the same classical trajectory by intercepting disjoint fragments of the environment---the
defining signature of quantum Darwinism. This indicates that the phase scrambling of
Feynman paths induced by repeated weak interactions in our formalism provides a microscopic,
trajectory-level mechanism underlying the proliferation of redundant classical
information central to quantum Darwinism. Under the conditions we have discussed---the system
being sufficiently macroscopic and the limit of weak measurements---information about the
classical paths is continuously mapped onto the probes which scatter and gives rise to
objectivity and redundancy, an essential ingredient for quantum
Darwinism~\cite{Brandao2015,Korbicz2021roadstoobjectivity,PhysRevA.91.032122,PhysRevLett.93.220401}.


Therefore, the derivation of this result, Eq.~\eqref{eq:central}, is another approach to
study quantum Darwinism and the emergence of objectivity, complementing the formulations
based on the decoherence paradigm. Quantum Darwinism has also been formulated for dynamical
systems at the level of the reduced density matrix, for instance the redundant recording of an
oscillator's pointer states in quantum Brownian motion~\cite{BlumeKohoutZurek2008}. The present
formulation instead works at the level of individual paths, the objective and redundantly encoded
entity being a single classical trajectory $\mathbf{r}_{cl}(\tau)$ selected as the fixed point of
probe-averaged phase scrambling. This helps us understand the quantum-to-classical
transition and the emergence of individual classical trajectories and objectivity across the
environment.

{A path-integral description of \emph{individual} continuously monitored trajectories is also
central to the stochastic path-integral (action) formalism of Chantasri, Dressel, and
Jordan~\cite{ChantasriDresselJordan2013,ChantasriJordan2015} and its continuous-variable
extension to a monitored oscillator~\cite{KarmakarLewalleJordan2022}, in which the joint
distribution of a measurement record and the conditioned trajectory is written as a phase-space
path integral and extremised to yield most-likely paths. Our construction shares that
record-conditioned, single-trajectory viewpoint but is aimed at a different question. That body of work deals with the statistics, optimal estimation, and feedback control of
the record of a single observer. Here we use the same continuous-monitoring kernel to ask
whether the recorded trajectory is objective, i.e. redundantly available to many independent
observers. The quantities of interest are therefore the redundancy, the pointer basis, and the
mutual-information plateau, and not the most likely path.
 A second distinction is dynamical: the
most-likely-path program treats the conditioned trajectory as the estimator to be optimized,
whereas we track the measurement back-action itself, which through the recoil-driven momentum
diffusion of Sec.~\ref{sec:window} sets a ceiling on how long a trajectory stays deterministic
before it is randomized into Brownian motion, and hence on how redundantly it can be objectively
recorded.}

{The many-observer side of this question has recently been sharpened in the pointer-state
setting: redundant environmental records enforce a \emph{consensus} among observers who read
independent fragments, so that those with sufficient information agree on the same pointer
state~\cite{TouilYanZurek2025}. We ask the analogous question for an individual recorded trajectory instead of a pointer
observable. In addition, we find a dynamical limit, the recoil ceiling derived above, on how
long such a consensus about a trajectory can persist before back-action randomizes it. This
limit has no counterpart in the ensemble picture.
}

\section{Discussion and Conclusions}

We have presented a path-integral formulation of the emergence of classical dynamics in a
quantum world by considering an approach that also incorporates continuous measurements. The
path-integral formulation provides the joint statistics of a sequence of measurements with each
Feynman path picking up an additional random phase due to measurements. The magnitude of this
phase is proportional to the measurement strength, and we give conditions under which the
dominant contribution to the probability amplitude comes from the trajectories in the vicinity
of the classical paths. The proliferation of this information across the environment, an
essential feature of quantum Darwinism, takes place via scattering of plane-wave probes by the
system. Extending to repeated measurements, we show that in the continuous limit, each system
trajectory picks up an additional phase due to the momentum kicks from the probes---the origin
of the back-action force. We provide conditions under which the measurement yields enough
``which-path'' information while keeping the wave packet sufficiently localized. This allows the
quantum-to-classical transition to be described at the level of individual measurement records.

We further show that the scattering which decoheres a trajectory also heats it, so that
decoherence and measurement back-action are related. This puts a bound on how redundantly an
individual classical trajectory can be recorded before back-action turns it into Brownian
motion; for a trapped particle the bound is set by the measurement resolution in units of the
zero-point motion. The bound is not restrictive for macroscopic systems. For a trapped particle
the redundancy reduces to a single record in the same limit in which the semiclassical
description of the orbit breaks down, while for a free particle the bound gives the time beyond
which the recorded trajectory is no longer deterministic, even though the classical action is
still large. The crossover involved is due to photon recoil and has already been observed in
levitated optomechanics. In our description it is the time after which a redundantly recorded
trajectory remains objective but is no longer deterministic.

Historically, the quantum-to-classical transition has been approached through the correspondence
principle~\cite{Shankar:102017,BOHR1928}, environment-induced
decoherence~\cite{zurek_1991,RevModPhys.75.715}, and the continuous measurement of quantum
systems~\cite{TSKquantumclassical,CavesMilburn}, as reviewed in the Introduction. The decoherence
program is naturally phrased at the level of the ensemble-averaged density matrix rather than the
single experiment; the path-integral formulation developed here is complementary to that picture: it yields the same
probability functional over measurement records but organises it around individual system
trajectories, so the distinction is one of emphasis rather than ontology.

Our approach rests on three ingredients: a sufficiently macroscopic system, weak continuous
measurements, and a scattering process that carries information to the environment. Relaxing the
first---taking a system far from the macroscopic limit, such as the $N$-slit
interferometer---returns the standard decoherence picture: a strong which-path measurement
scrambles the inter-path phases and leaves a density matrix diagonal in the pointer basis,
equivalent to the fluctuating-phase formulation of Stern, Aharonov and Imry~\cite{Stern,TanWalls}.
We give this analysis in Appendix~\ref{app:nslit}. The same scattering broadcasts the einselected
information throughout the environment, where it is accessible to many
observers~\cite{PhysRevLett.105.020404,Zurek2009}, as in quantum Darwinism.

It will be interesting to extend these ideas to study the quantum-to-classical transition and
the onset of quantum Darwinism for many-body systems. For the case of
a bipartite system of spin and harmonic oscillator~\cite{sghose}, the entanglement between the two causes
strong measurement back-action on the harmonic oscillator, and we cannot recover classical
trajectories in this system, regardless of the continuous measurements. In our framework, the
entanglement between constituent parts can be interpreted as a continuous mutual monitoring of
subsystems via a coupling interaction. Therefore, each subsystem acts as a probe for others
coupled to it. In such an extended, mutually monitored system the smooth weak-to-strong
measurement crossover found here for a single trajectory could sharpen into a genuine
measurement-induced phase transition, with the entanglement structure of the collective
state---rather than a single-particle suppression factor---acting as the order parameter that
separates a coherent quantum regime from one of redundant, objective classical records. The
degree to which each of these subsystems is macroscopic in comparison to
the entanglement strength will determine whether or not the collective motion can be described
as classical. It could be the case that a few degrees of freedom exhibit classical behavior
yielding classical trajectories while others behave in the ensemble sense. This will also help
us define the ``true'' classical limit of a many-body quantum system~\cite{PhysRevE.85.036208}.
As we have shown, the problem of taking the classical limit is much richer than merely taking
$\hbar\to0$ and involves an interplay of system coupling, degree of entanglement and the
relative actions of the subsystems. What kind of information then proliferates in the
environment in a Darwinian sense will be of significant interest to researchers working in
quantum simulations, quantum foundations and many-body physics. Finally, it is worth
mentioning the connections of our work to the emergence of many-body classical chaos from the
underlying quantum mechanics and the fundamental questions regarding the generation of
complexity and the emergence of classical Kolmogorov--Sinai entropy in many-body quantum
systems.

\begin{acknowledgments}
Harsh Arora and Bishal Kumar Das contributed equally to this work. We acknowledge Jess Riedel for many useful discussions. We thank Wojciech Hubert Zurek and Jarosław Korbicz
for pointing out useful literature. This work was
supported in part by grant DST/ICPS/QusT/Theme-3/2019/Q69, a New Faculty Seed Grant from IIT
Madras, and National Science Foundation grant PHY-2112890. The authors were further supported,
in part, by a grant from Mphasis to the Centre for Quantum Information, Communication, and
Computing (CQuICC) at IIT Madras and by the Prime Minister's Research Fellowship, Ministry of
Human Resource Development, Govt.\ of India. We also acknowledge funding by ANRF file number 
ANRF/ARGM/2025/002679/TS. We acknowledge the use of AI-based language tools, which assisted in improving the clarity and quality of the manuscript’s writing and editing. All scientific content and interpretations remain the authors’ own.

\end{acknowledgments}

\appendix

\section{Derivation of the measurement kernel}
\label{app:kernel}

In this appendix we provide in detail the steps leading to the central result
Eq.~\eqref{eq:central}. Considering a simultaneous measurement of the probes in the position
basis, the joint probability amplitude for the system's propagator and the measurement record
$\{\mathbf{r}_p(\tau),0\le\tau\le t\}$ follows from Eq.~\eqref{eq:Phi-joint}:
\begin{widetext}
\begin{equation}
\Phi(\mathbf{r}_p(t);\mathbf{r}_f,t_f,\mathbf{r}_i,0)=\lim_{M\to\infty}
\int\prod_{j=1}^{M+1}d\mathbf{K}_f^{(j)}S(\mathbf{K}_f^{(j)},\mathbf{K}_i)
\exp\!\left(i\!\int_0^t\!\Delta\mathbf{K}(\tau)\cdot\mathbf{r}_p(\tau)\right)Z(\mathbf{r}_f,t_f,\mathbf{r}_i,0).
\label{eq:A1}
\end{equation}
\end{widetext}
We begin with the following expression, in which only the variables being integrated over are
written explicitly:
\begin{equation}
\lim_{M\to\infty}\prod_{j=1}^{M+1}\int d\mathbf{K}_f^{(j)}S(\mathbf{K}_f^{(j)},\mathbf{K}_i)\,
e^{i\Delta\mathbf{K}(\tau_j)\cdot(\mathbf{r}_p(\tau_j)-\mathbf{r}_{cl}(\tau_j))}.
\label{eq:A2}
\end{equation}
To proceed, we work in polar coordinates, choosing the initial momentum of all probes along
the $z$-axis, $\mathbf{K}_i=|\mathbf{K}_i|\hat z$, and parameterising the final momenta by the
polar angle $\theta$ and azimuthal angle $\phi$:
\begin{equation}
\mathbf{K}_f=|\mathbf{K}_f|(\sin\theta\cos\phi\,\hat x+\sin\theta\sin\phi\,\hat y+\cos\theta\,\hat z).
\label{eq:A3}
\end{equation}
We assume that the system--probe interaction is weak, so the scattering can be considered
elastic, restricting the final momenta to lie on a sphere of radius $K_i$,
$|\mathbf{K}_f|=|\mathbf{K}_i|$. With the further assumption that the probes are equally likely
to scatter in all directions, the elastic scattering matrix takes the form quoted in the main
text, Eq.~\eqref{eq:Sansatz}, with the isotropic angular amplitude $A =\tfrac{1}{\sqrt{4\pi}}$; the $\sin\theta$ factor enters through the Jacobian $dK_f$. With this
restriction, the integrals over $\mathbf{K}_f^{(j)}$ can be carried out exactly. Expressing the
dot product $e^{i\Delta\mathbf{K}(\tau_j)\cdot(\mathbf{r}_p(\tau_j)-\mathbf{r}(\tau_j))}$ in
terms of polar coordinates:
\begin{widetext}
\begin{equation}
e^{i\Delta\mathbf{K}(\tau_j)\cdot(\mathbf{r}_p(\tau_j)-\mathbf{r}(\tau_j))}\equiv
\exp\!\left(i\frac{2\pi}{\lambda}\big[\sin\theta_j\cos\phi_j\,\delta r_1(\tau_j)
+\sin\theta_j\sin\phi_j\,\delta r_2(\tau_j)+\cos\theta_j\,\delta r_3(\tau_j)\big]\right)
e^{-i\frac{2\pi}{\lambda}\delta r_3(\tau_j)},
\label{eq:A5}
\end{equation}
\end{widetext}
where we have set $\delta\mathbf{r}(\tau_j)=\mathbf{r}_p(\tau_j)-\mathbf{r}_{cl}(\tau_j)$. Using
the Jacobian
$d\mathbf{K}_f^{(j)}\equiv|\mathbf{K}_f|^2\sin\theta_j\,d|\mathbf{K}_f^{(j)}|\,d\theta_j\,d\phi_j$
and performing the trivial modulus integral fixed by the elastic delta function, we are left
with the angular integrals
\begin{widetext}
\begin{align}
I_j&=\int_0^\pi d\theta_j\int_0^{2\pi}d\phi_j\,\sin\theta_j\,
\exp\!\left(i\frac{2\pi}{\lambda}\big[\sin\theta_j\cos\phi_j\,\delta r_1+\sin\theta_j\sin\phi_j\,\delta r_2+\cos\theta_j\,\delta r_3\big]\right)e^{-i\frac{2\pi}{\lambda}\delta r_3(\tau_j)}
\nonumber\\
&=4\pi\,e^{-i\frac{2\pi}{\lambda}\delta r_3(\tau_j)}\,
\frac{\sin\!\big(\frac{2\pi}{\lambda}|\mathbf{r}_p(\tau_j)-\mathbf{r}(\tau_j)|\big)}
{\frac{2\pi}{\lambda}|\mathbf{r}_p(\tau_j)-\mathbf{r}(\tau_j)|},
\label{eq:Ij}
\end{align}
\end{widetext}
where
$|\mathbf{r}_p(\tau_j)-\mathbf{r}(\tau_j)|=\sqrt{\delta r_1^2(\tau_j)+\delta r_2^2(\tau_j)+\delta r_3^2(\tau_j)}$~\cite{stein1971fourier}.
This is the sinc kernel quoted as Eq.~\eqref{eq:sinc}; it is strikingly similar to the
distribution of light after diffraction through a pinhole. Generalizing to the continuous limit
($M\to\infty$), the joint amplitude is
\begin{equation}
\Phi(\mathbf{r}_p(t);\mathbf{r}_f,t)=\left(\lim_{M\to\infty}\prod_{j=1}^{M+1}I_j\right)
\sqrt{\frac{\partial^2S_{cl}}{\partial r_b\,\partial r_a}}\,e^{\frac{iS_{cl}}{\hbar}}.
\label{eq:A7}
\end{equation}
Denoting $I=\lim_{M\to\infty}\prod_{j=1}^{M+1}I_j$ and working with the logarithm,
\begin{equation}
\log(I)=\lim_{M\to\infty}\sum_{j=1}^{M+1}\log(I_j).
\label{eq:A8}
\end{equation}
Using the product representation of the sinc function,
\begin{equation}
\frac{\sin(x)}{x}=\prod_{k=1}^{\infty}\left(1-\frac{x^2}{k^2\pi^2}\right),
\label{eq:B2}
\end{equation}
and expanding the logarithm with $\log(1-u)=-\sum_{m=1}^\infty u^m/m$, we obtain
\begin{align}
\log(I_j)&=-\sum_{k=1}^{\infty}\sum_{m=1}^{\infty}\frac{1}{m}\left(\frac{y_j}{k\pi}\right)^{2m}
+\log\!\left(4\pi\,e^{-i\frac{2\pi}{\lambda}\delta r_3(\tau_j)}\right),
\label{eq:B3}\\
\log(I)&=-\sum_{j}\sum_{k=1}^{\infty}\sum_{m=1}^{\infty}\frac{1}{m}\left(\frac{y_j}{k\pi}\right)^{2m},
\label{eq:B4}
\end{align}
where $y_j=\frac{2\pi}{\lambda}|\mathbf{r}_p(\tau_j)-\mathbf{r}(\tau_j)|$ and the
{residual phase factor, which does not affect the modulus, has been absorbed into a
normalization factor.} We may interchange the
sums over $m$ and $k$; the sum over $k$ is performed exactly, giving
$\sum_{k=1}^{\infty}k^{-2m}=\zeta(2m)$, where $\zeta$ denotes the Riemann zeta function.
Together with $(y_j/\pi)^{2m}=4^m|\delta\mathbf{r}|^{2m}/\lambda^{2m}$, this yields
\begin{widetext}
\begin{equation}
\log(I)=-\lim_{M\to\infty}\sum_{j=1}^{M+1}\sum_{m}\frac{4^m}{m}\,\zeta(2m)\,
\frac{\big[(\delta r_1(\tau_j))^2+(\delta r_2(\tau_j))^2+(\delta r_3(\tau_j))^2\big]^m}{\lambda^{2m}}.
\label{eq:B5}
\end{equation}
\end{widetext}
The limit $M\to\infty$ implies that the number of measurements also goes to infinity in the
interval $(0,t)$; hence each measurement must also become coarser. To take the
continuous limit we set $\lambda=\alpha/\sqrt{\Delta t}$, so that while the number of
measurements in the short time $\Delta t$ blows up, each measurement becomes coarser
($\lambda\to\infty$) in such a manner that $\lambda\sqrt{\Delta t}=\alpha$ converges as $\Delta t\to0$. With
this scaling the $m$-th term scales as $\Delta t^{\,m-1}$: only the $m=1$ term survives, the
leading correction ($m=2$) being $O(\Delta t)$ and all $m\ge2$ terms vanishing as
$\Delta t\to0$. Hence
\begin{align}
\log(I)&=-\frac{4\zeta(2)}{\alpha^2}\int_0^t d\tau\,\big\{(\delta r_1)^2+(\delta r_2)^2+(\delta r_3)^2\big\}
\nonumber\\
\Longrightarrow\ I&=\mathcal{N}\exp\!\left(-\frac{4\zeta(2)}{\alpha^2}\int_0^t d\tau\,|\delta\mathbf{r}|^2\right),
\label{eq:A10}
\end{align}
where $\mathcal{N}$ denotes an appropriate normalization constant. We arrive at the result
Eq.~\eqref{eq:central} in the main text by replacing the Riemann zeta function
$\zeta(2)=\pi^2/6$, which gives the suppression coefficient $2\pi^2/(3\alpha^2)$.

\section{Derivation of the branching structure}
\label{app:branching}

We derive Eq.~\eqref{eq:branch2} from the same scattering unitary
that produced Eq.~\eqref{eq:central}. Let the system start in a
superposition of $\mathcal N$ macroscopically distinct wave packets,
$|\Phi(0)\rangle=\sum_\alpha c_\alpha|\phi_\alpha\rangle$, whose
centroids follow candidate classical histories
$\mathbf r_\alpha(\tau)$ separated by at least $\Delta x$, with packet
width $w\ll\Delta x$ and $Kw\ll1$. Probes prepared in plane waves
$|\mathbf K_i\rangle$ scatter elastically at times $t_1<\dots<t_M$. A
single event acts on a position eigenstate of the system as
\begin{equation}
\hat U_S\,|\mathbf r\rangle|\mathbf K_i\rangle=|\mathbf r\rangle\otimes
|e(\mathbf r)\rangle,
\label{eq:event}
\end{equation}
with
\begin{equation}
|e(\mathbf r)\rangle=\int d\Omega\,A(\Omega)\,
e^{-i\Delta\mathbf K\cdot\mathbf r}\,|\mathbf K_f(\Omega)\rangle,
\label{eq:estate}
\end{equation}
where $\Delta\mathbf K=\mathbf K_f(\Omega)-\mathbf K_i$ and
$\langle e(\mathbf r)|e(\mathbf r)\rangle=1$ for the isotropic
normalization $A=1/\sqrt{4\pi}$; the momentum $-\hbar\Delta\mathbf K$
transferred to the system is the recoil analyzed in
Sec.~\ref{sec:window} and does not affect the argument here. Because
$Kw\ll1$, the phase $e^{-i\Delta\mathbf K\cdot\mathbf r}$ is constant
across a single packet to $O(Kw)$, so within each branch the packet
scatters as a point source at its centroid,
\begin{equation}
\hat U_S\,|\phi_\alpha\rangle|\mathbf K_i\rangle
=|\phi_\alpha\rangle\otimes|e(\mathbf r_\alpha(t_j))\rangle+O(Kw).
\end{equation}
Between events each branch propagates semiclassically, its centroid
following $\mathbf r_\alpha(\tau)$ and its internal phase supplying the
$e^{iS_{cl}^{(\alpha)}/\hbar}$ and van Vleck factors of
Eq.~\eqref{eq:central}. By linearity, after $M$ events
\begin{equation}
|\Psi(t)\rangle=\sum_\alpha c_\alpha\,|\psi_\alpha(t)\rangle
\bigotimes_{j=1}^{M}\big|e_j(\mathbf r_\alpha(t_j))\big\rangle,
\label{eq:branch-derived}
\end{equation}
up to an accumulated $O(MKw)$ correction, which is Eq.~\eqref{eq:branch2}. The conditional probe states overlap per event as
\begin{equation}
\langle e(\mathbf r)|e(\mathbf r')\rangle
=\!\int\! d\Omega\,|A|^{2}
e^{i\Delta\mathbf K\cdot(\mathbf r-\mathbf r')}
=\mathrm{sinc}\big(K|\mathbf r-\mathbf r'|\big)
\label{eq:eoverlap}
\end{equation}
for isotropic $A$---the same kernel that fixes the record noise,
Eq.~\eqref{eq:sigma2dt}, and the decoherence time,
Eq.~\eqref{eq:tauD}. The branching structure, the record statistics,
and the recoil therefore descend from one unitary. Cross-branch
coherences in any observable supported on the system are suppressed by
$\gamma_M=\prod_{j}\mathrm{sinc}(K\,\delta r_j)$, with
$\delta r_j=|\mathbf r_\alpha(t_j)-\mathbf r_\beta(t_j)|$
[Eq.~\eqref{eq:eoverlap}], and the decomposition is meaningful precisely
under the hierarchy $w\ll\Delta x$, $Kw\ll1$, $S_{cl}\gg\hbar$ assumed
throughout.

\section{Which-path measurement and the loss of interference in the $N$-slit experiment}
\label{app:nslit}

The loss of coherence, due to the elastic scattering of the environment states from the system,
leads to localization in phase space~\cite{FlemingPRA,Joos1985}. The theory of
decoherence due to an environment provides a phenomenological understanding of the
quantum--classical transition by suppressing superposition
states~\cite{schlosshauer2019quantumtoclassical,schlosshauer2007decoherence}. For an interferometry experiment involving
$N$ paths, the initial wave function of the system can be written as:
\begin{equation}
|\psi\rangle=\frac{1}{\sqrt N}\sum_{k=1}^{N}|\psi_k\rangle,
\label{eq:nslit-psi}
\end{equation}
where $|\psi_k\rangle$ represents the wave function at the beginning of the $k$th path. For
instance, in a grating with $N$ slits, each of these wave functions corresponds to a Gaussian
with mean at the centre of the $k$th slit and variance given by the slit width. The states
$|\psi_k\rangle$ are approximately orthogonal and distinguishable. The probability amplitude
for the system to be found at some position $x_s$ on the screen is simply found by
time-evolving the state and then taking overlaps:
\begin{equation}
\phi=\frac{1}{\sqrt N}\sum_{k=1}^{N}\langle x_s|\psi_k(t)\rangle=\frac{1}{\sqrt N}\sum_{k=1}^{N}\phi_k,
\label{eq:phi-amp}
\end{equation}
where each $\phi_k$ denotes the amplitude for the system to propagate from the $k$th slit to a
position on the screen or detector. In general, the time evolution leads to the states
$|\psi_k(t)\rangle$ being non-orthogonal, and hence the total probability at the screen is:
\begin{equation}
P=|\phi|^2=\frac{1}{N}\left(\sum_k|\phi_k|^2+2\sum_{i < j}\mathrm{Re}(\phi_i^*\phi_j)\right).
\label{eq:P-interf}
\end{equation}
The last term arises due to the interference between the paths. Under the semiclassical
approximation, $\phi_k=\sqrt{\frac{\partial^2S_{cl}}{\partial x_s\partial x_k}}\,e^{iS_{cl}/\hbar}$,
with $x_k$ denoting the position of the $k$th slit and $S_{cl}$ denoting the action of the
system evaluated along the classical path connecting the $k$th slit to the location
$x_s$~\cite{schulman2012techniques}. The partial derivative of the classical action provides the density
of paths that begin at the slits and end at $x_s$. Assuming equal weights for all the $\phi_k$,
we obtain:
\begin{equation}
P\sim1+\frac{2}{N}\sum_{i<j}\cos\!\left(\frac{S_{cl}^{(j)}-S_{cl}^{(i)}}{\hbar}\right).
\label{eq:P-cos}
\end{equation}
Therefore, the phase difference between the paths is simply the difference of the classical
actions. Since this difference is coherent, a stable interference pattern emerges from the $N$
slits. In this scenario, no information about the path that the system took is available.

In order to obtain which-path information, consider a measurement of the system's initial state
Eq.~\eqref{eq:nslit-psi}. This measurement leads to the following joint system--detector state:
\begin{equation}
|\psi_{SD}\rangle=\frac{1}{\sqrt N}\sum_{k=1}^{N}|\psi_k\rangle|D(x_k)\rangle.
\label{eq:psiSD}
\end{equation}
The states $|D(x_k)\rangle$ are normalized but not necessarily orthogonal. If the states
$|D(x_k)\rangle$ are perfectly orthogonal, then there exists a measurement basis which can
determine the state of the detector without ambiguity. This measurement then also determines
the state of the system. Then, since it is now determined which slit the system passes through,
interference fringes are completely lost. However, which-slit determination is not perfect if
the probe states have some overlap with each other. The interference pattern survives, although
it is diminished. The other extreme case occurs when the measurement provides no information at
all, with $|D(x_k)\rangle\equiv|D\rangle$ for all $k$.

The loss in fringe visibility, accompanied by a gain in which-slit information, is the
manifestation of quantum complementarity. Note that measuring the detector in a complementary
basis simply leads to a shift in the interference pattern. Let $\{|n\rangle\}$ denote some
complementary basis for the detector states, with $d$ the dimension of the detector Hilbert space. Expanding the states $|D(x_i)\rangle$ gives:
\begin{equation}
|D(x_k)\rangle = {\frac{1}{\sqrt{d}}\sum_{n=1}^{d}} e^{i\phi_n(x_k)}|n\rangle,
\label{eq:B6}
\end{equation}

with real phases $\phi_n(x_k)$; equal moduli $1/\sqrt{d}$ are the defining property of a basis complementary to the detector pointer states, and preserve the normalization $\langle D(x_k)|D(x_k)\rangle = 1$.
 The joint system--detector state is:
\begin{equation}
|\psi\rangle\equiv\frac{1}{\sqrt {Nd}}\sum_n\sum_{k=1}^{N}e^{i\phi_n(x_k)}|\psi_k\rangle|n\rangle.
\label{eq:psi-joint2}
\end{equation}
The conditional probability amplitude, assuming the measurement gives some state $|m\rangle$,
for the system to be found at position $x_s$ is:
\begin{equation}
\phi^{(m)}=\frac{1}{\sqrt {Nd}}\sum_k e^{i\phi_m(x_k)}\phi_k(t),
\label{eq:phi-m}
\end{equation}
where the $\phi_k$ correspond to the same probability amplitudes as in
Eq.~\eqref{eq:phi-amp}. The probability at the screen is simply $|\phi^{(m)}|^2$, and leads to an
interference pattern since $\phi^m$ is a sum over amplitudes corresponding to different paths.
However, compared to Eq.~\eqref{eq:phi-amp}, each of the $\phi_k$ has an additional phase. It is
straightforward to note that the minima and maxima are consequently shifted, and hence the
entire interference pattern shifts. Therefore, measurement in a complementary basis gives us
information about the relative phases ($\phi_m(x_k)$) of the system's wave
function~\cite{Zubairy2003PRA}, but gives no which-slit information. The local phase of
the wave function is related to the momentum distribution. Hence, changing the local phases
essentially acts to redirect the quantum system, similar to the refraction of
waves~\cite{Scully1991}. This local phase information is completely scrambled when
the measurement gives us which-slit information. Consider the overlap of $|\psi_{SD}\rangle$ in
Eq.~\eqref{eq:psiSD} with the joint state $|x_s\rangle|D(x_j)\rangle$ for some $j$:
\begin{widetext}
\begin{equation}
\langle {D(x_j)}|\langle x_s|\psi_{SD}\rangle = \frac{1}{\sqrt{N}}\sum_k \phi_k(t)\,{\frac{1}{d}}\sum_m e^{i[\phi_m(x_k) - \phi_m(x_j)]}.
\label{eq:overlap-Dj}
\end{equation}
\end{widetext}

The sum over $m$ gives the overlap $\langle D(x_j)|D(x_k)\rangle$. When the detector is read out in an orthonormal basis $\{|D(x_j)\rangle\}$---the strong-measurement configuration of interest below---the total probability at the
screen position $x_s$ is given by the sum of the modulus squared of the overlaps in
Eq.~\eqref{eq:overlap-Dj}, which adds up the probabilities corresponding to different
measurement outcomes:
\begin{widetext}
\begin{equation}
P=\sum_j|\langle D(x^{(j)})|\langle x_s|\psi\rangle|^2
=\frac{1}{N}\sum_j\left|\sum_k\phi_k(t)\frac{1}{d}\sum_m e^{i\phi_m(x_k)-i\phi_m(x^{(j)})}\right|^2.
\label{eq:P-sum}
\end{equation}
\end{widetext}
The amplitudes $\phi_k(t)$ can be replaced with the corresponding semiclassical propagators,
$\phi_k\sim e^{iS_{cl}/\hbar}$ (assuming equal weights for all $k$):
\begin{widetext}
\begin{equation}
P\sim\frac{1}{N}\sum_j\left|\sum_k\exp\!\left(\frac{iS_{cl}^{(k)}}{\hbar}\right) \frac{1}{d}
\sum_m e^{i\phi_m(x_k)-i\phi_m(x^{(j)})}\right|^2.
\label{eq:P-semiclass}
\end{equation}
\end{widetext}
The sum over $k$ corresponds to the sum of probability amplitudes, conditioned on the
measurement of the probe leading to state $|D(x_j)\rangle$. As compared to
Eq.~\eqref{eq:P-interf}, the sum over $m$ leads to complete scrambling of the phases when we
collect which-slit information. If the states $|D(x_j)\rangle$ are orthogonal
(corresponding to strong measurement), we find the sum
$\frac{1}{d}\sum_m e^{i\phi_m(x_k)-i\phi_m(x^{(j)})}=\delta_{jk}$. In this case, the probability is given
by:
\begin{equation}
P\sim\frac{1}{N}\sum_j|\phi_j|^2,
\label{eq:P-diagonal}
\end{equation}
where the sum over $j$ adds up the probability from each measurement outcome, since we are
averaging over all measurement outcomes. The strongest measurement will completely scramble
phases and hence result in Eq.~\eqref{eq:P-diagonal}~\cite{AntiZeno}. The outcome of
this experiment in the limit of strong measurements can be described by a density matrix with
only diagonal elements representing the probabilities $|\phi_j|^2$. Indeed, the loss of
coherence and interference patterns in various which-path experiments involving different
configurations of the double-slit setup can be understood in terms of randomization of the
wave function's phase~\cite{TanWalls,Scully1991}, as shown to be equivalent
to environmental dephasing~\cite{Stern}. Complete loss of coherence occurs
whenever the probability distribution governing this phase difference is slowly varying over
the interval $[0,2\pi]$~\cite{Bertet2001}.

Therefore, decoherence and the emergence of classicality are manifested as the vanishing of the
off-diagonal elements of the density matrix which transforms quantum systems with genuine
quantum correlations into mathematical entities that can be treated as classical probability
distributions. While the above analysis is in terms of the footprint of the system on the
detector states, it has been shown by Stern, Aharonov and Imry~\cite{Stern}
that a formulation in terms of the fluctuating phase difference between the multiple
interfering paths is completely equivalent and leads to loss of coherence~\cite{TanWalls}
and complementarity. The process of measurement, which involves interactions with external
agents, alters the phases associated with each of the interfering paths. Further, the
scattering process shows that information about the einselected system proliferates throughout
the environment and is readily accessible to observers~\cite{PhysRevLett.105.020404,Zurek2009}.

\section{Derivation of the $t^3$ back-action law}
\label{app:backaction}
{Here we derive Eq.~\eqref{eq:BAwander}. The recoil kicks act as a white-noise force on the
system momentum, so the momentum performs a random walk with diffusion coefficient $D_p$ of
Eq.~\eqref{eq:Dp},
\begin{equation}
\langle\Delta p^2(t)\rangle=2D_p\,t,\qquad
\langle\Delta p(t')\,\Delta p(t'')\rangle=2D_p\,\min(t',t'') .
\label{eq:pdiff}
\end{equation}
For an inertial (free) coordinate the velocity is $\dot{\mathbf r}=\mathbf p/m$, so the centroid
displacement is the time integral of this diffusing momentum,
$\Delta x_{\rm BA}(t)=\tfrac{1}{m}\int_0^t\Delta p(t')\,dt'$. Squaring and averaging with
Eq.~\eqref{eq:pdiff},
\begin{align}
\langle\Delta x_{\rm BA}^2(t)\rangle
&=\frac{1}{m^2}\int_0^t\!\!\int_0^t\langle\Delta p(t')\,\Delta p(t'')\rangle\,dt'\,dt''
\nonumber\\
&=\frac{2D_p}{m^2}\int_0^t\!\!\int_0^t\min(t',t'')\,dt'\,dt'' .
\label{eq:xba-double}
\end{align}
The double integral over the square $[0,t]^2$ is
\begin{equation}
\int_0^t\!\!\int_0^t\min(t',t'')\,dt'\,dt''
=2\int_0^t\!dt'\int_0^{t'}\!t''\,dt''
=\frac{t^3}{3},
\label{eq:mintt}
\end{equation}
which gives $\langle\Delta x_{\rm BA}^2(t)\rangle=\tfrac{2D_p}{3m^2}\,t^3$, and substituting the
decoherence--diffusion relation $D_p=\hbar^2/(\tau_D\Delta x^2)$ of Eq.~\eqref{eq:FDT} yields the
second form in Eq.~\eqref{eq:BAwander}. 

The growth is cubic in time, and not linear as for a coordinate that diffuses directly, because
it is the momentum that diffuses and the position is its time integral. The same law governs
photon-recoil heating of a free levitated particle ~\cite{Jain2016}.
 The result assumes a constant $D_p$ and white momentum noise
uncorrelated with the localization; relaxing these fixes only the $O(1)$ prefactor, which is why
Eq.~\eqref{eq:BAwander} is written with ``$\simeq$''. For the harmonic trap the same momentum
diffusion instead raises the oscillator energy linearly, giving
$\langle\Delta x_{\rm BA}^2(t)\rangle=D_p t/m^2\omega^2$ as used in the main text.}

\bibliography{ref1_clean}

@book{haake1991quantum,
  author        = {Haake, Fritz},
  title         = {Quantum Signatures of Chaos},
  year          = {1991},
  publisher     = {Springer}
}

@article{asher_peres_chaos,
  author        = {Peres, Asher},
  title         = {Chaotic evolution in quantum mechanics},
  journal       = {Phys. Rev. E},
  volume        = {53},
  pages         = {4524--4527},
  year          = {1996},
  doi           = {10.1103/PhysRevE.53.4524}
}

@book{peres1997quantum,
  author        = {Peres, Asher},
  title         = {Quantum Theory: Concepts and Methods},
  series        = {Fundamental Theories of Physics},
  volume        = {57},
  year          = {1997},
  publisher     = {Springer},
  doi           = {10.1007/0-306-47120-5}
}

@article{koopman_pnas.17.5.315,
  author        = {Koopman, B. O.},
  title         = {{H}amiltonian Systems and Transformation in {H}ilbert Space},
  journal       = {Proc. Natl. Acad. Sci. U.S.A.},
  volume        = {17},
  pages         = {315--318},
  year          = {1931},
  doi           = {10.1073/pnas.17.5.315}
}

@article{von_neuman_1932,
  author        = {von Neumann, J.},
  title         = {Zur {O}peratorenmethode in der klassischen {M}echanik},
  journal       = {Ann. Math.},
  volume        = {33},
  pages         = {587--642},
  year          = {1932},
  doi           = {10.2307/1968537}
}

@article{stepping_stone_hamiltonian,
  author        = {Jordan, Thomas F.},
  title         = {Steppingstones in {H}amiltonian dynamics},
  journal       = {Am. J. Phys.},
  volume        = {72},
  pages         = {1095--1099},
  year          = {2004},
  doi           = {10.1119/1.1737394}
}

@article{TanmoyPRL,
  author        = {Bhattacharya, Tanmoy and Habib, Salman and Jacobs, Kurt},
  title         = {Continuous Quantum Measurement and the Emergence of Classical Chaos},
  journal       = {Phys. Rev. Lett.},
  volume        = {85},
  pages         = {4852--4855},
  year          = {2000},
  doi           = {10.1103/PhysRevLett.85.4852}
}

@article{Berry2002,
  author        = {Berry, Michael},
  title         = {Singular limits},
  journal       = {Phys. Today},
  volume        = {55},
  pages         = {10--11},
  year          = {2002},
  doi           = {10.1063/1.1485555}
}

@article{EPR,
  author        = {Einstein, A. and Podolsky, B. and Rosen, N.},
  title         = {Can Quantum-Mechanical Description of Physical Reality Be Considered Complete?},
  journal       = {Phys. Rev.},
  volume        = {47},
  pages         = {777--780},
  year          = {1935},
  doi           = {10.1103/PhysRev.47.777}
}

@article{JSB,
  author        = {Bell, J. S.},
  title         = {On the {E}instein {P}odolsky {R}osen paradox},
  journal       = {Physics Physique Fizika},
  volume        = {1},
  pages         = {195--200},
  year          = {1964},
  doi           = {10.1103/PhysicsPhysiqueFizika.1.195}
}

@article{bohr_1935,
  author        = {Bohr, N.},
  title         = {Can Quantum-Mechanical Description of Physical Reality be Considered Complete?},
  journal       = {Phys. Rev.},
  volume        = {48},
  pages         = {696--702},
  year          = {1935},
  doi           = {10.1103/PhysRev.48.696}
}

@incollection{bohreinstein,
  author        = {Bohr, Niels},
  title         = {Discussion with {E}instein on epistemological problems in atomic physics},
  booktitle     = {Niels Bohr Collected Works},
  volume        = {7},
  pages         = {339--381},
  year          = {1996},
  publisher     = {Elsevier},
  doi           = {10.1016/S1876-0503(08)70379-7}
}

@article{zurek_1991,
  author        = {Zurek, Wojciech H.},
  title         = {Decoherence and the Transition from Quantum to Classical},
  journal       = {Phys. Today},
  volume        = {44},
  pages         = {36--44},
  year          = {1991},
  doi           = {10.1063/1.881293}
}

@article{BermanZaslavsky1978,
  author        = {Berman, G. P. and Zaslavsky, G. M.},
  title         = {Condition of stochasticity in quantum nonlinear systems},
  journal       = {Physica A},
  volume        = {91},
  pages         = {450--460},
  year          = {1978},
  doi           = {10.1016/0378-4371(78)90190-5}
}

@article{ZurekPaz1994,
  author        = {Zurek, W. H. and Paz, J. P.},
  title         = {Decoherence, chaos, and the second law},
  journal       = {Phys. Rev. Lett.},
  volume        = {72},
  pages         = {2508},
  year          = {1994},
  doi           = {10.1103/PhysRevLett.72.2508}
}

@article{HabibShizumeZurek1998,
  author        = {Habib, Salman and Shizume, Kosuke and Zurek, Wojciech Hubert},
  title         = {Decoherence, chaos, and the correspondence principle},
  journal       = {Phys. Rev. Lett.},
  volume        = {80},
  pages         = {4361--4365},
  year          = {1998},
  doi           = {10.1103/PhysRevLett.80.4361}
}

@article{Mensky1979,
  author        = {Mensky, M. B.},
  title         = {Quantum restrictions for continuous observation of an oscillator},
  journal       = {Phys. Rev. D},
  volume        = {20},
  pages         = {384--387},
  year          = {1979},
  doi           = {10.1103/PhysRevD.20.384}
}

@book{Mensky1993,
  author        = {Mensky, Michael B.},
  title         = {Continuous Quantum Measurements and Path Integrals},
  year          = {1993},
  publisher     = {IOP Publishing},
  address       = {Bristol}
}

@article{MenskyOnofrioPresilla1991,
  author        = {Mensky, M. B. and Onofrio, R. and Presilla, C.},
  title         = {Continuous quantum monitoring of position of nonlinear oscillators},
  journal       = {Phys. Lett. A},
  volume        = {161},
  pages         = {236--240},
  year          = {1991},
  doi           = {10.1016/0375-9601(91)90010-6}
}

@misc{Mensky2002RPI,
  author        = {Mensky, M. B.},
  title         = {Quantum continuous measurements, dynamical role of information and restricted path integrals},
  howpublished  = {arXiv:quant-ph/0212112},
  year          = {2002}
}

@article{Caves1,
  author        = {Caves, Carlton M.},
  title         = {Quantum mechanics of measurements distributed in time. {A} path-integral formulation},
  journal       = {Phys. Rev. D},
  volume        = {33},
  pages         = {1643--1665},
  year          = {1986},
  doi           = {10.1103/PhysRevD.33.1643}
}

@article{Caves2,
  author        = {Caves, Carlton M.},
  title         = {Quantum mechanics of measurements distributed in time. {II}. {C}onnections among formulations},
  journal       = {Phys. Rev. D},
  volume        = {35},
  pages         = {1815--1830},
  year          = {1987},
  doi           = {10.1103/PhysRevD.35.1815}
}

@book{taylor2012scattering,
  author        = {Taylor, J. R.},
  title         = {Scattering Theory: The Quantum Theory of Nonrelativistic Collisions},
  series        = {Dover Books on Engineering},
  year          = {2012},
  publisher     = {Dover Publications}
}

@article{TanWalls,
  author        = {Tan, S. M. and Walls, D. F.},
  title         = {Loss of coherence in interferometry},
  journal       = {Phys. Rev. A},
  volume        = {47},
  pages         = {4663--4676},
  year          = {1993},
  doi           = {10.1103/PhysRevA.47.4663}
}

@article{Storey1994,
  author        = {Storey, Pippa and Tan, Sze and Collett, Matthew and Walls, Daniel},
  title         = {Path detection and the uncertainty principle},
  journal       = {Nature},
  volume        = {367},
  pages         = {626--628},
  year          = {1994},
  doi           = {10.1038/367626a0}
}

@book{Heller2018,
  author        = {Heller, Eric J.},
  title         = {The Semiclassical Way to Dynamics and Spectroscopy},
  year          = {2018},
  publisher     = {Princeton University Press},
  address       = {Princeton},
  doi           = {10.23943/9781400890293}
}

@article{dewittmorette,
  author        = {DeWitt-Morette, C{\'e}cile},
  title         = {The semiclassical expansion},
  journal       = {Ann. Phys. (N.Y.)},
  volume        = {97},
  pages         = {367--399},
  year          = {1976},
  doi           = {10.1016/0003-4916(76)90041-5}
}

@book{Mensky2000book,
  author        = {Mensky, M. B.},
  title         = {Quantum Measurements and Decoherence: Models and Phenomenology},
  year          = {2000},
  publisher     = {Kluwer Academic},
  address       = {Dordrecht},
  doi           = {10.1007/978-94-015-9566-7}
}

@article{PresillaOnofrioTambini,
  author        = {Presilla, C. and Onofrio, R. and Tambini, U.},
  title         = {Measurement quantum mechanics and experiments on quantum {Z}eno effect},
  journal       = {Ann. Phys. (N.Y.)},
  volume        = {248},
  pages         = {95--121},
  year          = {1996},
  doi           = {10.1006/aphy.1996.0052}
}

@article{Zurek2009,
  author        = {Zurek, Wojciech Hubert},
  title         = {Quantum {D}arwinism},
  journal       = {Nat. Phys.},
  volume        = {5},
  pages         = {181--188},
  year          = {2009},
  doi           = {10.1038/nphys1202}
}

@article{e24111520,
  author        = {Zurek, Wojciech Hubert},
  title         = {Quantum Theory of the Classical: Einselection, Envariance, Quantum {D}arwinism and Extantons},
  journal       = {Entropy},
  volume        = {24},
  pages         = {1520},
  year          = {2022},
  doi           = {10.3390/e24111520}
}

@article{PhysRevA.73.062310,
  author        = {Blume-Kohout, Robin and Zurek, Wojciech H.},
  title         = {Quantum {D}arwinism: Entanglement, branches, and the emergent classicality of redundantly stored quantum information},
  journal       = {Phys. Rev. A},
  volume        = {73},
  pages         = {062310},
  year          = {2006},
  doi           = {10.1103/PhysRevA.73.062310}
}

@article{LeeKorbicz2024,
  author        = {Lee, Tae-Hun and Korbicz, Jaros{\l{}}aw K.},
  title         = {Complementarity between decoherence and information retrieval from the environment},
  journal       = {Phys. Rev. A},
  volume        = {109},
  pages         = {032221},
  year          = {2024},
  doi           = {10.1103/PhysRevA.109.032221}
}

@article{PhysRevLett.105.020404,
  author        = {Riedel, C. Jess and Zurek, Wojciech H.},
  title         = {Quantum {D}arwinism in an Everyday Environment: Huge Redundancy in Scattered Photons},
  journal       = {Phys. Rev. Lett.},
  volume        = {105},
  pages         = {020404},
  year          = {2010},
  doi           = {10.1103/PhysRevLett.105.020404}
}

@book{BohrenHuffman1983,
  author        = {Bohren, Craig F. and Huffman, Donald R.},
  title         = {Absorption and Scattering of Light by Small Particles},
  year          = {1983},
  publisher     = {Wiley},
  address       = {New York},
  doi           = {10.1002/9783527618156}
}

@article{6773024,
  author        = {Shannon, C. E.},
  title         = {A mathematical theory of communication},
  journal       = {Bell Syst. Tech. J.},
  volume        = {27},
  pages         = {379--423},
  year          = {1948},
  doi           = {10.1002/j.1538-7305.1948.tb01338.x}
}

@article{TuziemskiKorbicz2015,
  author        = {Tuziemski, J. and Korbicz, J. K.},
  title         = {Dynamical objectivity in quantum {B}rownian motion},
  journal       = {Europhys. Lett.},
  volume        = {112},
  pages         = {40008},
  year          = {2015},
  doi           = {10.1209/0295-5075/112/40008}
}

@article{RiedelZurekZwolak2012,
  author        = {Riedel, C. Jess and Zurek, Wojciech H. and Zwolak, Michael},
  title         = {The rise and fall of redundancy in decoherence and quantum {D}arwinism},
  journal       = {New J. Phys.},
  volume        = {14},
  pages         = {083010},
  year          = {2012},
  doi           = {10.1088/1367-2630/14/8/083010}
}

@article{ZwolakQuanZurek2009,
  author        = {Zwolak, Michael and Quan, H. T. and Zurek, Wojciech H.},
  title         = {Quantum {D}arwinism in a mixed environment},
  journal       = {Phys. Rev. Lett.},
  volume        = {103},
  pages         = {110402},
  year          = {2009},
  doi           = {10.1103/PhysRevLett.103.110402}
}

@article{ZwolakQuanZurek2010,
  author        = {Zwolak, Michael and Quan, H. T. and Zurek, Wojciech H.},
  title         = {Redundant imprinting of information in nonideal environments: Objective reality via a noisy channel},
  journal       = {Phys. Rev. A},
  volume        = {81},
  pages         = {062110},
  year          = {2010},
  doi           = {10.1103/PhysRevA.81.062110}
}

@article{ZwolakRiedelZurek2014,
  author        = {Zwolak, Michael and Riedel, C. Jess and Zurek, Wojciech H.},
  title         = {Amplification, Redundancy, and the Quantum {C}hernoff Information},
  journal       = {Phys. Rev. Lett.},
  volume        = {112},
  pages         = {140406},
  year          = {2014},
  doi           = {10.1103/PhysRevLett.112.140406}
}

@article{Jain2016,
  author        = {Jain, Vijay and Gieseler, Jan and Moritz, Clemens and Dellago, Christoph and Quidant, Romain and Novotny, Lukas},
  title         = {Direct Measurement of Photon Recoil from a Levitated Nanoparticle},
  journal       = {Phys. Rev. Lett.},
  volume        = {116},
  pages         = {243601},
  year          = {2016},
  doi           = {10.1103/PhysRevLett.116.243601}
}

@article{GonzalezBallestero2021,
  author        = {Gonzalez-Ballestero, C. and Aspelmeyer, M. and Novotny, L. and Quidant, R. and Romero-Isart, O.},
  title         = {Levitodynamics: Levitation and control of microscopic objects in vacuum},
  journal       = {Science},
  volume        = {374},
  pages         = {eabg3027},
  year          = {2021},
  doi           = {10.1126/science.abg3027}
}

@article{Brandao2015,
  author        = {Brand{\~a}o, Fernando G. S. L. and Piani, Marco and Horodecki, Pawe{\l}},
  title         = {Generic emergence of classical features in quantum {D}arwinism},
  journal       = {Nat. Commun.},
  volume        = {6},
  pages         = {7908},
  year          = {2015},
  doi           = {10.1038/ncomms8908}
}

@article{Korbicz2021roadstoobjectivity,
  author        = {Korbicz, J. K.},
  title         = {Roads to objectivity: {Q}uantum {D}arwinism, {S}pectrum {B}roadcast {S}tructures, and {S}trong quantum {D}arwinism -- a review},
  journal       = {Quantum},
  volume        = {5},
  pages         = {571},
  year          = {2021},
  doi           = {10.22331/q-2021-11-08-571}
}

@article{PhysRevA.91.032122,
  author        = {Horodecki, R. and Korbicz, J. K. and Horodecki, P.},
  title         = {Quantum origins of objectivity},
  journal       = {Phys. Rev. A},
  volume        = {91},
  pages         = {032122},
  year          = {2015},
  doi           = {10.1103/PhysRevA.91.032122}
}

@article{PhysRevLett.93.220401,
  author        = {Ollivier, Harold and Poulin, David and Zurek, Wojciech H.},
  title         = {Objective Properties from Subjective Quantum States: Environment as a Witness},
  journal       = {Phys. Rev. Lett.},
  volume        = {93},
  pages         = {220401},
  year          = {2004},
  doi           = {10.1103/PhysRevLett.93.220401}
}

@article{BlumeKohoutZurek2008,
  author        = {Blume-Kohout, Robin and Zurek, Wojciech H.},
  title         = {Quantum {D}arwinism in Quantum {B}rownian Motion},
  journal       = {Phys. Rev. Lett.},
  volume        = {101},
  pages         = {240405},
  year          = {2008},
  doi           = {10.1103/PhysRevLett.101.240405}
}

@article{ChantasriDresselJordan2013,
  author        = {Chantasri, A. and Dressel, J. and Jordan, A. N.},
  title         = {Action principle for continuous quantum measurement},
  journal       = {Phys. Rev. A},
  volume        = {88},
  pages         = {042110},
  year          = {2013},
  doi           = {10.1103/PhysRevA.88.042110}
}

@article{ChantasriJordan2015,
  author        = {Chantasri, A. and Jordan, A. N.},
  title         = {Stochastic path-integral formalism for continuous quantum measurement},
  journal       = {Phys. Rev. A},
  volume        = {92},
  pages         = {032125},
  year          = {2015},
  doi           = {10.1103/PhysRevA.92.032125}
}

@article{KarmakarLewalleJordan2022,
  author        = {Karmakar, Tathagata and Lewalle, Philippe and Jordan, Andrew N.},
  title         = {Stochastic Path-Integral Analysis of the Continuously Monitored Quantum Harmonic Oscillator},
  journal       = {PRX Quantum},
  volume        = {3},
  pages         = {010327},
  year          = {2022},
  doi           = {10.1103/PRXQuantum.3.010327}
}

@misc{TouilYanZurek2025,
  author        = {Touil, Akram and Yan, Bin and Zurek, Wojciech H.},
  title         = {Consensus about classical reality in a quantum universe},
  howpublished  = {arXiv:2503.14791},
  year          = {2025},
  doi           = {10.48550/arXiv.2503.14791}
}

@book{Shankar:102017,
  author        = {Shankar, Ramamurti},
  title         = {Principles of Quantum Mechanics},
  year          = {1980},
  publisher     = {Plenum},
  address       = {New York, NY}
}

@article{BOHR1928,
  author        = {Bohr, N.},
  title         = {The quantum postulate and the recent development of atomic theory},
  journal       = {Nature},
  volume        = {121},
  pages         = {580--590},
  year          = {1928},
  doi           = {10.1038/121580a0}
}

@article{RevModPhys.75.715,
  author        = {Zurek, Wojciech Hubert},
  title         = {Decoherence, einselection, and the quantum origins of the classical},
  journal       = {Rev. Mod. Phys.},
  volume        = {75},
  pages         = {715--775},
  year          = {2003},
  doi           = {10.1103/RevModPhys.75.715}
}

@article{TSKquantumclassical,
  author        = {Bhattacharya, Tanmoy and Habib, Salman and Jacobs, Kurt},
  title         = {Continuous quantum measurement and the quantum to classical transition},
  journal       = {Phys. Rev. A},
  volume        = {67},
  pages         = {042103},
  year          = {2003},
  doi           = {10.1103/PhysRevA.67.042103}
}

@article{CavesMilburn,
  author        = {Caves, Carlton M. and Milburn, G. J.},
  title         = {Quantum-mechanical model for continuous position measurements},
  journal       = {Phys. Rev. A},
  volume        = {36},
  pages         = {5543--5555},
  year          = {1987},
  doi           = {10.1103/PhysRevA.36.5543}
}

@article{Stern,
  author        = {Stern, Ady and Aharonov, Yakir and Imry, Yoseph},
  title         = {Phase uncertainty and loss of interference: A general picture},
  journal       = {Phys. Rev. A},
  volume        = {41},
  pages         = {3436--3448},
  year          = {1990},
  doi           = {10.1103/PhysRevA.41.3436}
}

@article{sghose,
  author        = {Ghose, Shohini and Alsing, Paul and Deutsch, Ivan and Bhattacharya, Tanmoy and Habib, Salman},
  title         = {Transition to classical chaos in a coupled quantum system through continuous measurement},
  journal       = {Phys. Rev. A},
  volume        = {69},
  pages         = {052116},
  year          = {2004},
  doi           = {10.1103/PhysRevA.69.052116}
}

@article{PhysRevE.85.036208,
  author        = {Casati, Giulio and Guarneri, Italo and Reslen, Jose},
  title         = {Classical dynamics of quantum entanglement},
  journal       = {Phys. Rev. E},
  volume        = {85},
  pages         = {036208},
  year          = {2012},
  doi           = {10.1103/PhysRevE.85.036208}
}

@book{stein1971fourier,
  author        = {Elias M. Stein and Guido Weiss},
  title         = {Introduction to Fourier Analysis on Euclidean Spaces},
  year          = {1971},
  publisher     = {Princeton University Press},
  address       = {Princeton},
  doi           = {10.1515/9781400883899},
  note          = {See formula for uniform Fourier transform over a sphere, p. 154}
}

@article{FlemingPRA,
  author        = {Gallis, Michael R. and Fleming, Gordon N.},
  title         = {Environmental and spontaneous localization},
  journal       = {Phys. Rev. A},
  volume        = {42},
  pages         = {38--48},
  year          = {1990},
  doi           = {10.1103/PhysRevA.42.38}
}

@article{Joos1985,
  author        = {Joos, E. and Zeh, H. D.},
  title         = {The emergence of classical properties through interaction with the environment},
  journal       = {Z. Phys. B},
  volume        = {59},
  pages         = {223--243},
  year          = {1985},
  doi           = {10.1007/BF01725541}
}

@misc{schlosshauer2019quantumtoclassical,
  author        = {Schlosshauer, Maximilian},
  title         = {The quantum-to-classical transition and decoherence},
  howpublished  = {arXiv:1404.2635},
  year          = {2019}
}

@book{schlosshauer2007decoherence,
  author        = {Schlosshauer, Maximilian},
  title         = {Decoherence and the Quantum-to-Classical Transition},
  series        = {The Frontiers Collection},
  year          = {2007},
  publisher     = {Springer},
  doi           = {10.1007/978-3-540-35775-9}
}

@book{schulman2012techniques,
  author        = {Schulman, L. S.},
  title         = {Techniques and Applications of Path Integration},
  series        = {Dover Books on Physics},
  year          = {2012},
  publisher     = {Dover Publications}
}

@article{Zubairy2003PRA,
  author        = {Kapale, Kishore T. and Qamar, Shahid and Zubairy, M. Suhail},
  title         = {Spectroscopic measurement of an atomic wave function},
  journal       = {Phys. Rev. A},
  volume        = {67},
  pages         = {023805},
  year          = {2003},
  doi           = {10.1103/PhysRevA.67.023805}
}

@article{Scully1991,
  author        = {Scully, Marian O. and Englert, Berthold-Georg and Walther, Herbert},
  title         = {Quantum optical tests of complementarity},
  journal       = {Nature},
  volume        = {351},
  pages         = {111--116},
  year          = {1991},
  doi           = {10.1038/351111a0}
}

@article{AntiZeno,
  author        = {Kaulakys, B. and Gontis, V.},
  title         = {Quantum anti-{Z}eno effect},
  journal       = {Phys. Rev. A},
  volume        = {56},
  pages         = {1131--1137},
  year          = {1997},
  doi           = {10.1103/PhysRevA.56.1131}
}

@article{Bertet2001,
  author        = {Bertet, P. and Osnaghi, S. and Rauschenbeutel, A. and Nogues, G. and Auffeves, A. and Brune, M. and Raimond, J. M. and Haroche, S.},
  title         = {A complementarity experiment with an interferometer at the quantum--classical boundary},
  journal       = {Nature},
  volume        = {411},
  pages         = {166--170},
  year          = {2001},
  doi           = {10.1038/35075517}
}

\end{document}